\documentclass[aps,preprint,showpacs,amsmath,amssymb]{revtex4}
\usepackage{epsfig}
\usepackage{wasysym}
\usepackage{amssymb}

\begin{document}

\newcommand{\bvec}[1]{\mbox{\boldmath ${#1}$}}
\title{Electromagnetic production of kaon near threshold}
\author{T. Mart}
\affiliation{Departemen Fisika, FMIPA, Universitas Indonesia, Depok 16424, 
  Indonesia}
\date{\today}
\begin{abstract}
Kaon photo- and electroproduction 
off a proton near the production threshold 
are investigated by utilizing 
an isobar model. The background amplitude of the model is constructed
from Feynman diagrams, whereas the resonance
term is calculated by using the multipole formalism.
It is found that both pseudoscalar and pseudovector 
models can nicely describe the available 
photoproduction data up to $W=50$ MeV above the threshold.
The $\Lambda$ resonance $S_{01}(1800)$ is found to play
an important role in improving the model. 
In the case of double polarization observables $O_x$ and $O_z$
our result corroborates the finding of Sandorfi {\it et al.}
Due to the large contributions of the $K^*$ and $K_1$ vector 
mesons, extending the model 
to the case of electroproduction is almost impossible 
unless either special form factors that strongly suppress their
contributions are introduced or all 
hadronic coupling constants are
refitted to both photo- and electroproduction databases,
simultaneously. It is also 
concluded that investigation of the kaon electromagnetic form
factor is not recommended near the threshold region.
\end{abstract}

\maketitle

\section{Introduction}
\label{sec:intro}
It has been known that a comprehensive phenomenological 
analysis of kaon photoproduction is still far away from what 
we expected one or two decades ago, in spite of the fact that 
there are ample experimental
data with good quality recently provided by modern accelerators 
such as JLab in Newport News, ELSA in Bonn, and SPRING8
in Osaka. The reason for this set back is 
because of the high threshold 
energy, which not only ``switches on'' the strangeness degree
of freedom, but also increases the level of complexity 
of phenomenological or theoretical models that attempt 
to describe the process. Already close to the production 
threshold ($W_{\rm thr.}=1609$ MeV for $K^+\Lambda$ photoproduction), 
a number of established nucleon resonances contribute
to the process, for instance the $S_{11}(1650)$, $D_{15}(1675)$, 
and $F_{15}(1680)$. Even below the threshold there are the
well known 
Roper resonance $P_{11}(1440)$ along with two other four-star
resonances $D_{13}(1520)$ and $S_{11}(1535)$. Furthermore,
the presence of strangeness degree of freedom insists on 
the use of SU(3) symmetry in the reaction mechanism, rather
than SU(2) as in the case of pion production, which clearly makes 
the theoretical formalism more difficult.

In the last decades tremendous efforts have been devoted to
model kaon photoproduction process. Most of them have been 
performed in the framework of tree-level isobar models  
\cite{abw,Adelseck:1990ch,williams,Han:1999ck,kaon-maid,
Janssen:2001wk,delaPuente:2008bw}, coupled channel approaches  
\cite{Feuster:1998cj,Chiang:2001pw,Shklyar:2005xg,Julia-Diaz:2006is,
Anisovich:2007bq}, Regge model \cite{Guidal:1997hy}, 
or quark models \cite{Li:1995sia,Lu:1995bk}. Extending 
the energy range of isobar models to higher region 
by utilizing Regge formalism 
has also been pursued recently 
\cite{Mart:2003yb,Mart:2004au,Corthals:2005ce,Vancraeyveld:2009qt}.
At present, experimental data are abundant from threshold up to
2.5 GeV. In this energy range there are 19 nucleon resonances
in the Particle Data Book \cite{Amsler:2009} table (PDG), 
which may propagate in the $s$-channel intermediate states. In
addition, there are hyperon and meson resonances that may 
also influence
the background terms of the process. To take into account all these
excited states is naturally a daunting task, since their hadronic
and electromagnetic coupling constants are mostly unknown. There is a 
possibility to fit all these unknown parameters, however, the
accuracies of presently available data still 
allow for too many possible solutions.
Furthermore, the extracted couplings are quite often 
``nonphysical'', in the sense that their values are
much smaller or much larger than the widely known pion
coupling constant $g_{\pi NN}$. In the literature, a number of recipes 
have been put forward to avoid this problem, e.g., by limiting 
the number of resonances, by taking into account as many as 
possible constraints that are relevant to the process, or by 
combining these two methods.

Except limiting the energy of interest, there is no firm argument
for limiting the number of resonances that should be put into the model. 
In the
previous works, the goal to construct a very simple elementary
operator for use in nuclear physics was sometimes used for
this purpose \cite{abw,williams,kaon-maid}. The argument of duality is
also proposed to this end \cite{saghai96}. As a result, the number
of phenomenological models describing kaon photoproduction
was quickly increasing in the last decades. These models 
differ chiefly by the number and type of resonances used as 
well as, quite obviously, the values of extracted coupling 
constants.

The exact values of coupling constants are a long-standing
problem. Even for the main coupling constants $g_{K\Lambda N}$ 
and $g_{K\Sigma N}$ there is so far no consensus as to which values
should be used in the electromagnetic production of kaon. 
The SU(3) symmetry dictates the values of 
$g_{K\Lambda N}/\sqrt{4\pi} = -4.4 {\rm ~to~} -3.0$ and 
$g_{K\Sigma N}/\sqrt{4\pi} = +0.9  {\rm ~to~} +1.3$, 
given that the SU(3) is broken at the level of 20\%
\cite{Adelseck:1990ch}. These values are consistent with 
those extracted from the $K$-$N$ or $Y$-$N$ 
scattering data \cite{anto,boz}. However, the QCD sum rules
predicted significantly smaller values, i.e. 
$g_{K\Lambda N}/\sqrt{4\pi}=-1.96$ and 
$g_{K\Sigma N}/\sqrt{4\pi}=+0.33$ \cite{choe}. On the other hand,
the values extracted from kaon photoproduction are mostly much 
smaller than the SU(3) prediction, unless hadronic form factors 
were used in hadronic vertices or some 
absorptive factors are applied to reduce the Born terms \cite{tkb}.
These facts indicate
that the problem of the $g_{K\Lambda N}$ and $g_{K\Sigma N}$
values is far from settled at present. Other coupling constants, such as
those of $K^*$ and $K_1$ intermediate states can be also estimated with
the help of SU(3) symmetry and a number of parameters extracted
from other reactions. Their values are by all means less accurate
than the values of $g_{K\Lambda N}$ and $g_{K\Sigma N}$.

A quick glance to the $K^+\Lambda$ photoproduction data base will
reveal that there are more than 100 data points of differential
cross section for the energy range from threshold up to 50 MeV
above the threshold, available from the SAPHIR \cite{Glander:2003jw}
and CLAS \cite{Bradford:2005pt} collaborations. This indicates
that a phenomenological analysis of kaon photoproduction near the 
production threshold is already possible. To the best of our knowledge,
there was no such an analysis performed with these data. The latest
study of kaon photoproduction from threshold up to 
14 MeV above the $E_{\gamma,{\rm thr}}^{\rm lab}$
was performed by Cheoun
{\it et al.} \cite{Cheoun:1996kn} more than a decade ago. Since there
were no data available at that time, Cheoun {\it et al.} predicted
the total cross section of the $\gamma + p\to {K^+}+\Lambda$ process
by varying the value of $g_{K\Lambda N}/\sqrt{4\pi}$ from 0 to 4
and studied the difference between the pseudoscalar (PS) and
pseudovector (PV) couplings in this reaction. They argued that
measurement of the total cross section can determine the real
value of this coupling constant. We note that for
the SU(3) value of the $g_{K\Lambda N}/\sqrt{4\pi}$, the 
total cross sections for both PS and PV couplings 
at 10 MeV above  $E_{\gamma,{\rm thr}}^{\rm lab}$ are predicted to be 
around 5 $\mu$b, whereas for the value accepted by the
QCD sum rule the cross sections would be slightly more than 
1 $\mu$b. It is interesting to compare these predictions 
with the currently available data (see Fig.~\ref{fig:total} 
in Section \ref{sec:result-photo}) and find 
that none of these predictions is right, since at this energy
the experimental cross section is found to be less than 0.4 $\mu$b.

From the facts presented above it is clear that investigation 
of kaon photoproduction near the production threshold is very 
important, because it could provide very important information
on the simplest form of the reaction mechanism as well as information 
on the background terms. In the isobar model this also means
information on the $t$- and $u$-channel intermediate
states that contribute to the
background terms. Such information is obviously very
difficult to obtain at high energies due to the complicated 
structure of reaction amplitude at this stage. 
Furthermore, since models 
that describe the production at threshold are in general 
quite simple, the individual contributions of intermediate 
states to this process can be easily studied. In summary,
the result of this investigation should become a stepping
stone for the construction of extended models describing
the photo- and electroproduction reaction at higher energies. 

In this paper we present our analysis on the 
$\gamma + p\to {K^+}+\Lambda$ process near the production threshold,
i.e. up to 50 MeV above the $W_{\rm thr.}$, as well its extension
to the electroproduction case $e + p\to e+K^++\Lambda$. 
We note that a new version of kaon photoproduction data
from CLAS collaboration has been published 
recently \cite{McCracken:2009ra}. However, since the behavior of 
these data at the threshold region is similar to that shown by the
previous ones \cite{Bradford:2005pt}, we believe that it is 
sufficient to use the previous version of
CLAS data \cite{Bradford:2005pt} 
for the photoproduction process in the following
discussion for the sake of simplicity.
As a starting point 
we consider the resonances suggested by Cheoun {\it et al.} 
\cite{Cheoun:1996kn} (see Table II of Ref.~\cite{Cheoun:1996kn}) 
for the possible resonance intermediate states in 
the energy range of interest. However, 
different from the work of Cheoun {\it et al.}, 
to simplify the fitting process in this work 
we fix the values of the $g_{K\Lambda N}$ and $g_{K\Sigma N}$ to the
SU(3) symmetry prediction, as in the case of Kaon-Maid 
\cite{kaon-maid}, though we do not use the hadronic form
factors in all hadronic vertices. For the resonance formalism, we
use the Breit-Wigner multipole form as suggested by Drechsel {\it et al.} 
\cite{hanstein99} and Tiator {\it et al.} \cite{Tiator:2003uu}.
The advantage of using this formalism as compared to the covariant
one is that it provides a direct comparison of the extracted 
helicity photon couplings with the PDG values. Moreover, the
multipole formalism does not produce unnecessary additional 
background terms that will interfere with the pure ones, 
as in the case of the covariant calculation. 
To further reduce the number of free parameters in our model we also fix
the resonance parameters to the PDG values. 

This paper is organized as follows. In Section \ref{sec:ps} we present
the formalism of our model in the PS theory. 
Section \ref{sec:resonance} briefly discusses resonance formalism
of our model. In Section \ref{sec:pv} we present the difference between
PS and PV amplitudes. The numerical results and comparison between
experimental data for kaon photoproduction 
and model calculations will be given in Section
\ref{sec:result-photo}. In Section \ref{sec:electroproduction}
we discuss the extension of our model to the case of
electroproduction and 
the effect of the available kaon electroproduction data 
on the result of our previous fit. 
In Section \ref{sec:conclusion} we summarize our findings.

\section{Pseudoscalar coupling}
\label{sec:ps}
\begin{figure}[!t]
  \begin{center}
    \leavevmode
    \epsfig{figure=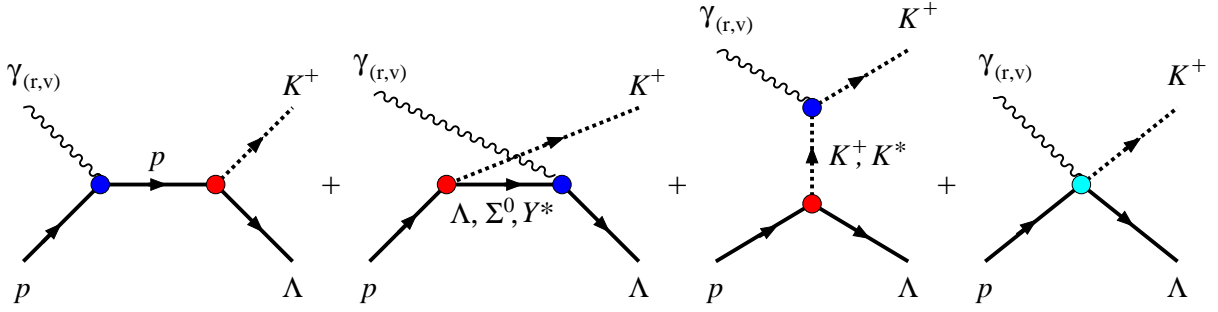,width=160mm}
    \caption{(Color online) Feynman diagrams of the background
	amplitude. Note that the last diagram (contact
        or seagull diagram) is used to maintain gauge invariance 
        in the pseudovector theory.}
   \label{fig:feynman} 
  \end{center}
\end{figure}

Let us consider photo- and electroproduction process that can
be represented as a real and virtual photon production
\begin{eqnarray}
  \gamma_{\rm r,v}(k) + p (p_p) \to K^+(q_K)+\Lambda (p_\Lambda)~.
\end{eqnarray}
As discussed in the previous section, the background terms  
of this process are obtained from a series of tree-level Feynman 
diagrams shown in Fig.~\ref{fig:feynman}. 
They consist of the standard $s$-, $u$-, and $t$-channel Born terms 
along with the $K^{*+}(892)$ and $K_1(1270)$ $t$-channel vector meson. 
The energy near threshold can be considered as low energy, 
therefore, we would 
expect that no hadronic form factors is required in all hadronic 
vertices of the diagrams in Fig.~\ref{fig:feynman}. This has
an obvious advantage, i.e., we can further limit the number of 
free parameters as well as uncertainties in our model. In 
Ref.~\cite{Bydzovsky:2006wy} it is shown that the use of 
hadronic form factor could lead to an underprediction of 
differential cross section data at forward angles. 

Using the standard procedure in the pseudoscalar theory 
the transition amplitude for kaon photo- and electroproduction 
off a proton reads
\begin{eqnarray}
  \label{eq:ps}
  {\cal M}^{\rm ps} &=& {\bar u}_\Lambda (\bvec{p}_\Lambda) \biggl[ ig_{K\Lambda N} \gamma_5
  \Bigl\{ \frac{p\!\!\!/_p+k\!\!\!/+m_p}{s-m_p^2}\Bigl( \epsilon\!\!/
  eF_1^p +i\sigma^{\mu\nu}\epsilon_{\mu}k_{\nu} \mu_p F_2^p \Bigr)
  -\frac{k\cdot\epsilon}{k^2} eF_1^p \Bigr\} 
  \nonumber\\ & &
  + i\sigma^{\mu\nu}\epsilon_{\mu}k_{\nu} \mu_\Lambda F_2^\Lambda 
  \frac{p\!\!\!/_\Lambda -k\!\!\!/+m_\Lambda}{u-m_\Lambda^2} 
  ig_{K\Lambda N}\gamma_5 
  + ig_{K\Lambda N}\gamma_5 \Bigl\{ \frac{(2q_K-k)\cdot\epsilon}{t-m_K^2}+
  \frac{k\cdot\epsilon}{k^2} \Bigr\} eF^K
  \nonumber\\ & &
  +\frac{i}{M(t-m_{K^*}^2+im_{K^*}\Gamma_{K^*})} 
  \Bigl\{ g^V_{K^*\Lambda N}\gamma_\mu-
  \frac{g^T_{K^*\Lambda N}}{m_p+m_\Lambda}
  i\sigma^{\mu\nu}(q_K-k)_\nu \Bigr\}
  \nonumber\\ & &
  \times i\varepsilon_{\mu\nu\rho\sigma}
  \epsilon^\nu k^\rho q_K^\sigma\, g_{K^*K\gamma} F^{K^*}
  \nonumber\\ & &
  +\frac{1}{M(t-m_{K_1}^2+im_{K_1}\Gamma_{K_1})}\Bigl\{ 
  g^V_{K_1\Lambda N}\gamma^\mu\gamma_5+\frac{g^T_{K_1\Lambda N}}{m_p+m_\Lambda}
  (p\!\!\!/_\Lambda-p\!\!\!/_p)\gamma^\mu\gamma_5\Bigr\}
  \nonumber\\ & &
  \times\left\{ (q_K-k)\cdot\epsilon\, k_\mu 
    - (q_K-k)\cdot k\,\epsilon_\mu\right\}\, g_{K_1K\gamma}
  F^{K_1}
  \nonumber\\ & &
  + i\sigma^{\mu\nu}\epsilon_\mu k_\nu \mu_T F_2^T 
  \frac{p\!\!\!/_{\Sigma} -k\!\!\!/+m_{\Sigma}}{u-m_{\Sigma}^2} 
  ig_{K\Sigma N}\gamma_5
  \nonumber\\ & &
  + i\sigma^{\mu\nu}\epsilon_\mu k_\nu \mu_{Y^*} F_2^{Y^*} 
  \frac{p\!\!\!/_{Y^*} -k\!\!\!/+m_{Y^*}}{u-m_{Y^*}^2+im_{Y^*}\Gamma_{Y^*}} 
  ig_{KY^* N}\gamma_5
  \,\biggr]\, u_p({\bvec p}_p) ~,
\end{eqnarray}
where  $\mu_{p}$ and $\mu_{\Lambda}$ denote the magnetic moments of 
the proton and $\Lambda$-hyperon, respectively, 
$\epsilon$ is the photon polarization vector, 
$F^i$ are the electromagnetic
form factors of the $i$ intermediate states that will be
discussed in Section~\ref{sec:electroproduction},
and $M=1$ GeV is introduced
to make the coupling constants of $K^*$ and $K_1$ dimensionless. 

The transition amplitude in Eq.~(\ref{eq:ps}) can be written as 
\cite{deo} 
\begin{eqnarray}
M_{\rm fi} & = & \bar u(\bvec{p}_{\Lambda}) 
\sum_{j=1}^6 A_{j}(s,t,k^2) M_{j} u(\bvec{p}_{p})~ ,
\label{trans}
\end{eqnarray}
where $s$ and $t$ are the usual Mandelstam variables, defined by
\begin{eqnarray}
s = (k+p_p)^2 ,~~ t = (k-q_K)^2 ,~~ u ~=~ (k-p_\Lambda)^2 ~, 
\label{mandelstam}
\end{eqnarray} 
whereas the gauge and Lorentz invariant matrices $M_{j}$ are given by
\begin{subequations}
\begin{eqnarray}
M_{1} & = & {\textstyle \frac{1}{2}} \gamma_{5} \left( \epsilon \!\!/ k 
\!\!\!/ - k \!\!\!/ \epsilon \!\!/ \right)~ ,\\
M_{2} & = & \gamma_{5}\left[ (2q_K-k) \cdot \epsilon P \cdot k - (2q_K-k) 
\cdot k P \cdot \epsilon \right]~ ,\\
\label{sixm}
M_{3} & = & \gamma_{5} \left( q_K\cdot k \epsilon \!\!/ - q_K\cdot \epsilon k
\!\!\!/ \right) ~ ,\\
M_{4} & = & i \epsilon_{\mu \nu \rho \sigma} \gamma^{\mu} q_K^{\nu}
\epsilon^{\rho} k^{\sigma}~ ,\\
M_{5} & = & \gamma_{5} \left( q_K\cdot \epsilon k^{2} - q_K\cdot k k \cdot 
\epsilon \right) ~ ,\\
M_{6} & = & \gamma_{5} \left( k \cdot \epsilon k \!\!\!/ - k^{2} \epsilon \!\!/
\right)~ ,
\end{eqnarray}
\end{subequations}
where $P = \frac{1}{2}(p_{p} + p_{\Lambda})$, and $\epsilon_{\mu \nu \rho
\sigma}$ is the four dimensional Levi-Civita tensor with
$\epsilon_{0123} = +1$. 
We note that the definition of the invariant matrices above is
slightly different from those given in Refs.~\cite{denn,pasquini},
specifically the $M_4$ amplitude. 
The functions $A_i$ therefore read
\begin{subequations}
\label{eq:born_ps}
\begin{eqnarray}
A_{1}^{\rm PS} & = & -\frac{e g_{K \Lambda N}}{s - m_{p}^{2}} \Bigl(
F_{1}^{p} + 
\kappa_{p} \frac{m_{p} - m_{\Lambda}}{2 m_{p}}F_{2}^{p} \Bigr) 
- \frac{e g_{K \Lambda N}}{u - m_{\Lambda}^{2}} 
\frac{m_{\Lambda} - m_{p}}{2 m_{\Lambda}} \kappa_{\Lambda} F_{2}^{\Lambda}
 \nonumber\\
& & -  \frac{e G_{K \Sigma N}}{u - m_{\Sigma}^{2}} 
\frac{m_{\Sigma} - m_{p}}{m_{\Sigma} + m_{\Lambda}} F_{2}^{T} 
- \frac{G^T_{K^*}tF^{K^*}}{M(t-m_{K^*}^2+im_{K^*}\Gamma_{K^*})(m_p + m_\Lambda)}
 \nonumber\\
& & +\frac{eG_{Y^*}F_2^{Y^*}}{u-m_{Y^*}^2+im_{Y^*}\Gamma_{Y^*}}
\left\{-\frac{m_p+m_{Y^*}}{m_\Lambda+m_{Y^*}}+\frac{i\Gamma_{Y^*}}{
2(m_\Lambda+m_{Y^*})}\right\}
~ ,\\
A_{2}^{\rm PS} & = & \frac{e g_{K \Lambda N}}{s - m_{p}^{2}}  
\frac{2 F_{1}^{p}}{t - m_{K}^{2}} 
+ \frac{G^T_{K^*}F^{K^*}}{M(t-m_{K^*}^2+im_{K^*}\Gamma_{K^*})(m_p+ m_\Lambda)}
\left(1+\frac{k^2}{t-m_K^2}\right)
\nonumber\\
& &- \frac{G^T_{K_1}F^{K_1}}{M(t-m_{K_1}^2+im_{K_1}\Gamma_{K_1})(m_p+ m_\Lambda)}
\left(1+\frac{k^2}{t-m_K^2}\right)
~ ,\\
A_{3}^{\rm PS} & = & \frac{e g_{K \Lambda N}}{s - m_{p}^{2}} 
\frac{\kappa_{p} F_{2}^{p}}{2 m_{p}} - 
\frac{e g_{K \Lambda N}}{u - m_{\Lambda}^{2}}  \frac{\kappa_{\Lambda}
F_{2}^{\Lambda}}{2 m_{\Lambda}} 
-  \frac{e G_{K \Sigma N}}{u - m_{\Sigma}^{2}} 
 \frac{F_{2}^{T}}{m_{\Sigma} + m_{\Lambda}}\nonumber\\&& 
- \frac{G^T_{K^*}F^{K^*}}{M(t-m_{K^*}^2+im_{K^*}\Gamma_{K^*})}
\frac{m_\Lambda - m_p}{m_\Lambda + m_p}\nonumber\\&& 
+ \frac{(m_\Lambda +m_p)G^V_{K_1}+ 
(m_\Lambda -m_p)G^T_{K_{1}}}{M(t-m_{K_1}^2+im_{K_1}\Gamma_{K_1})}
\frac{F^{K_1}}{m_\Lambda + m_p}
 \nonumber\\
& & +\frac{eG_{Y^*}F_2^{Y^*}}{u-m_{Y^*}^2+im_{Y^*}\Gamma_{Y^*}}
\frac{1}{m_\Lambda+m_{Y^*}}
~ ,\\
A_{4}^{\rm PS} & = & \frac{e g_{K \Lambda N}}{s - m_{p}^{2}} 
\frac{\kappa_{p} F_{2}^{p}}{2 m_{p}} + 
\frac{e g_{K \Lambda N}}{u - m_{\Lambda}^{2}}  \frac{\kappa_{\Lambda}
F_{2}^{\Lambda}}{2 m_{\Lambda}} 
+ \frac{e G_{K \Sigma N}}{u - m_{\Sigma}^{2}} 
 \frac{F_{2}^{T}}{m_{\Sigma} + m_{\Lambda}}
\nonumber\\&& 
+ \frac{G^V_{K^*}F^{K^*}}{M(t-m_{K^*}^2+im_{K^*}\Gamma_{K^*})}
-\frac{eG_{Y^*}F_2^{Y^*}}{u-m_{Y^*}^2+im_{Y^*}\Gamma_{Y^*}}
\frac{1}{m_\Lambda+m_{Y^*}}
~ ,\\
A_{5}^{\rm PS} & = & \frac{e g_{K \Lambda N}}{t - m_{K}^{2}}  \left\{
\left(-\frac{2}{k^2}
 + \frac{1}{s - m_{p}^{2}}\right) F_{1}^{p}
+ \frac{2}{k^{2}} F^{K}\right\}
\label{eq:born_ps5}
\nonumber\\
& & - \frac{G^T_{K^*}F^{K^*}}{M(t-m_{K^*}^2+im_{K^*}\Gamma_{K^*})(m_p+ m_\Lambda)}
\left\{ \frac{(s - m_{p}^{2}) - (u-m_{\Lambda}^{2})}{2(t - m_{K}^{2})}\right\}
\nonumber\\
& & + \frac{G^T_{K_1}F^{K_1}}{M(t-m_{K_1}^2+im_{K_1}\Gamma_{K_1})(m_p+ m_\Lambda)}
\left\{ \frac{(s - m_{p}^{2}) - (u-m_{\Lambda}^{2})}{2(t - m_{K}^{2})}\right\}
~ ,\\
A_{6}^{\rm PS} & = & - \frac{G^T_{K^*}F^{K^*}}{M(t-m_{K^*}^2+im_{K^*}\Gamma_{K^*})}
\frac{m_\Lambda - m_p}{m_\Lambda + m_p}\nonumber\\&& 
+ \frac{(m_\Lambda +m_p)G^V_{K_1}+ 
(m_\Lambda -m_p)G^T_{K_{1}}}{M(t-m_{K_1}^2+im_{K_1}\Gamma_{K_1})}
\frac{F^{K_1}}{m_\Lambda + m_p}
~ ,
\end{eqnarray}
\end{subequations}
where $\kappa_{p}$ and $\kappa_{\Lambda}$ denote the anomalous 
magnetic moments of the proton and $\Lambda$ hyperon; 
$G^V_{K^*} = g_{K^* K\gamma}g_{K^* YN}^V$, 
$G^T_{K^*} = g_{K^* K\gamma}g_{K^* YN}^T$,
$G^V_{K_1} = g_{K_1 K\gamma}g_{K_1 YN}^V$,
$G^T_{K_1} = g_{K_1 K\gamma}g_{K_1 YN}^T$,
$G_{Y^*} = \kappa_{Y^*}g_{KY^*N}$, 
and $G_{K \Sigma N} = \kappa_{T} g_{K \Sigma N}$. 
Note that the gauge invariance ensures that $A_5$ is still
finite at the photon point $k^2=0$. Thus, there is no singularity
in Eq.~(\ref{eq:born_ps5}) in the limit of $k\to 0$ and the 
transition from electroproduction to photoproduction is still smooth.

\section{The Resonance Amplitude}
\label{sec:resonance}
Since experimental data used in this calculation is limited up to
50 MeV above the threshold, only the resonance state $S_{11}(1650)$ 
contributes to the process. As a consequence only the 
resonant electric multipole $E_{0+}$ 
amplitude exists and the resonance contribution is extremely simplified.
Following the work of Drechsel {\it et al.} \cite{hanstein99}
and Tiator {\it et al.} \cite{Tiator:2003uu} 
(see also the discussion in Section II.B of Ref.~\cite{Mart:2006dk}),
we can write the resonant electric multipole in the Breit-Wigner 
form as
\begin{eqnarray}
  \label{eq:em_multipole}
  E_{0+}(W) &=& {\bar E}_{0+} \, c_{K\Lambda }\, \frac{f_{\gamma R}(W)\, 
    \Gamma_{\rm tot}(W) m_R\, f_{K R}(W)}{m_R^2-W^2-im_R\Gamma_{\rm tot}(W)}~ e^{i\phi} ~,
  \label{eq:m_multipole}
\end{eqnarray}
where $W$ represents the total c.m. energy, the isospin factor 
$c_{K\Lambda }=-1$  \cite{Mart:2006dk}, 
$f_{KR}$ is the usual
Breit-Wigner factor describing the decay of a resonance $R$ with a total width
$\Gamma_{\rm tot}(W)$ and physical mass $m_R$. The $f_{\gamma R}$ indicates
the $\gamma NR$ vertex and $\phi$ represents the phase angle. 
The Breit-Wigner factor $f_{KR}$ is given by 
\begin{eqnarray}
  \label{eq:f_KR}
  f_{KR}(W) &=& \left[\frac{1}{2\pi}\frac{k_W}{|\bvec{q}_K|}\frac{m_p}{W}
  \frac{\Gamma_{K\Lambda }}{\Gamma_{\rm tot}^2}\right]^{1/2}~~,~~~~
  k_W ~=~ \frac{W^2-m_p^2}{2W}~,
\end{eqnarray}
with $m_p$ the proton mass. The energy dependent partial 
width $\Gamma_{K\Lambda }$ is defined through
\begin{eqnarray}
  \label{eq:Gamma_KL}
  \Gamma_{K\Lambda } &=& \beta_K\Gamma_R \frac{|\bvec{q}_K|}{q_R}
  \,\frac{W_R}{W}~,
\end{eqnarray}
where $\beta_K$ is the single kaon branching ratio, 
$\Gamma_R$ and $q_R$ are the total width and kaon 
c.m. momentum at $W=m_R$, respectively. 
The $\gamma NR$ vertex is parameterized through
\begin{eqnarray}
  \label{eq:f_gammaR}
  f_{\gamma R} &=& \frac{k_W}{k_R} ~,
\end{eqnarray}
where $k_R$ is equal to $k_W$ calculated at $W=m_R$. 

The total width appearing in Eqs.~(\ref{eq:m_multipole}) 
and (\ref{eq:f_KR}) is the sum of $\Gamma_K$ and the 
``inelastic'' width $\Gamma_{\rm in}$. In this work
we assume the dominance of the pion decay channel and we 
parameterize the width by using
\begin{eqnarray}
  \label{eq:Gamma_tot}
  \Gamma_{\rm tot} ~=~ \Gamma_{K\Lambda }+\Gamma_{\rm in}~,~~~~
  \Gamma_{\rm in} ~=~ (1-\beta_K) \Gamma_R \left(\frac{q_\pi}{q_0}
  \right)^{4} \left(\frac{X^2+q_0^2}{X^2+q_\pi^2}\right)^{2}~,
\end{eqnarray}
with $q_\pi$ the momentum of the $\pi$ in the decay of $R\to\pi+N$ 
in c.m. system, $q_0=q_\pi$ calculated at $W=m_R$, and the 
damping parameter $X$ is assumed to be 500 MeV \cite{hanstein99}.

The electric multipole photon coupling ${\bar E}_{0+}$ 
in Eq.\,(\ref{eq:em_multipole}) is related to the helicity photon coupling
$A_{1/2}^{0+}$ through ${\bar E}_{0+}=-A_{1/2}^{0+}$.
The scalar multipole photon coupling ${\bar S}_{0+}$ has the
same form as Eq.~(\ref{eq:em_multipole}). It is related to the
helicity photon coupling $S_{1/2}^{0+}$ through 
${\bar S}_{0+}=-\sqrt{2}S_{1/2}^{0+}$. To calculate the cross
section we combine the CGLN amplitudes from the background
and resonance terms. The CGLN amplitudes 
$F_1$-$F_6$  \cite{germar} for the background
terms are calculated from the functions $A_1$-$A_6$ 
given in Eq.~(\ref{eq:born_ps}), in the case of PS coupling. 
Since we have only $S_{11}(1650)$ as the resonance,
the CGLN amplitudes in the resonance part becomes quite simple, i.e.,
\begin{subequations}
  \label{eq:cgln_res}
  \begin{eqnarray}
    F_1 &=& {\bar E}_{0+} ~,\\
    F_5 &=& \frac{k_0}{|\bvec{k}|} {\bar S}_{0+} ~.
  \end{eqnarray}
\end{subequations}

\section{Pseudovector coupling}
\label{sec:pv}
In the pseudovector theory we have to change the vertex $\gamma_{5}
g_{K\Lambda N}$ to $\gamma_{5} q\!\!\!/_K g_{K\Lambda N}^{\rm PV}$, where $q_K$ 
is the momentum of the kaon leaving the vertex and the pseudovector
coupling constant $g_{K\Lambda N}^{\rm PV}$ is related to the pseudoscalar
one through $g_{K\Lambda N}^{\rm PV}=g_{K\Lambda N}/(m_p+m_\Lambda)$
\cite{deo}. 
To maintain gauge invariance, the so-called 
contact (seagull) term shown in Fig.~\ref{fig:feynman} is added. 
Therefore, the functions $A_i$ in the pseudovector theory are 
related to those of the pseudoscalar theory 
given by Eq.~(\ref{eq:born_ps}) through
\begin{subequations}
\begin{eqnarray}
\label{eq:born_pv1}
A_{1}^{\rm PV} & = & A_{1}^{\rm PS}- \lambda_{\rm PV}
\left[\frac{e g_{K\Lambda N}}{m_{p}+ 
m_{\Lambda}}\left(\frac{\kappa_{p} F_{2}^{p}}{2 m_{p}} +
\frac{\kappa_{\Lambda} F_{2}^{\Lambda}}{2 m_{\Lambda}}\right) -
\frac{e G_{K\Sigma N}}{m_{\Lambda} + m_{\Sigma}}~ 
\frac{ F_{2}^{T}}{m_{p} + m_{\Sigma}} \nonumber \right. \\ &&
-\left. \frac{e G_{Y^*} 
F_{2}^{Y^{*}}}{u - m_{Y^{*}}^{2} + 
i m_{Y^{*}} \Gamma_{Y^{*}}}~ \frac{1}{m_{\Lambda}+m_{Y^{*}}}
 \biggl\{ 2m_{Y^*} + \frac{u - m_{Y^{*}}^{2}}{m_{Y^{*}} + 
m_{p}} - \frac{i\Gamma_{Y^{*}}}{2} \biggr\}\right]
 ,~~\\
A_{3}^{\rm PV} & = & A_{3}^{{\rm PS}} - \lambda_{\rm PV}\frac{e G_{Y^*} 
F_{2}^{Y^{*}}}{u - m_{Y^{*}}^{2} + 
i m_{Y^{*}} \Gamma_{Y^{*}}}~\frac{1}{m_{p}+m_{Y^{*}}}
 \biggl\{ \frac{2m_{Y^{*}}}{m_{\Lambda} + m_{Y^{*}}} -  
\frac{i\Gamma_{Y^{*}}}{2(m_{\Lambda} + m_{Y^{*}})} 
\biggr\} ,~~\\
A_{4}^{\rm PV} & = & A_{4}^{{\rm PS}} + \lambda_{\rm PV}\frac{e G_{Y^*} 
F_{2}^{Y^{*}}}{u - m_{Y^{*}}^{2} + i m_{Y^{*}} \Gamma_{Y^{*}}}~ 
\frac{1}{m_{p}+m_{Y^{*}}}\biggl\{ \frac{2m_{Y^{*}}}{m_{\Lambda}+m_{Y^{*}}} 
-\frac{i\Gamma_{Y^{*}}}{2(m_{\Lambda} 
+ m_{Y^{*}})} \biggr\} ,~~\\
A_{6}^{\rm PV} & = & A_{6}^{\rm PS}- \lambda_{\rm PV}
\frac{e g_{K\Lambda N}}{m_{p} + 
m_{\Lambda}} \left( \frac{F_{1}^{p}}{k^{2}} - \frac{F^{c}}{k^{2}}\right)  ,
\label{eq:born_pv6}
\end{eqnarray}
\end{subequations}
where $\lambda_{\rm PV}$ represents the PS-PV mixing parameter, i.e.,
$\lambda_{\rm PV}=1$ for the pure PV coupling,
$\lambda_{\rm PV}=0$ for the pure PS coupling, and
$0< \lambda_{\rm PV}< 1$ for the model with a PS-PV mixed coupling.
The impact of the contact term is obvious in 
Eq.~(\ref{eq:born_pv6}), since without this term
the longitudinal function $A_6$ would become singular at the photon
point $k^2=0$.

\section{Results for Photoproduction}
\label{sec:result-photo}
The cross sections and polarization observables are calculated
from the function  $A_i$ described above by using the standard formulas
given in Ref. \cite{germar}. These calculated observables are fitted 
to experimental data by adjusting the unknown coupling constants 
using the CERN-MINUIT code.

\begin{table}[!t]
  \centering
  \caption{Properties of the particles \cite{Amsler:2009} 
    that might contribute to this process. This list is based on the work 
    reported in Ref. \cite{Cheoun:1996kn}. The particles used in the 
    present analysis are indicated by $\surd$ in the last column.}
  \label{tab:particle_used}
  \begin{ruledtabular}
  \begin{tabular}[c]{ccccccc}
    Short & Resonance & I & $J^\pi$ & Mass & Width & 
     Used in the\\[-1ex]
    Notation & Notation & & &(MeV) &(MeV) & Present Work\\
    \hline
    $p$ &&$\frac{1}{2}$&$\frac{1}{2}^+$& $938.272$ & - & $\surd$ \\
    $K^+$ &&$\frac{1}{2}$&$0^-$& $493.677$ & - & $\surd$ \\
    $\Lambda$ &&0&$\frac{1}{2}^+$& $1115.683$ & - & $\surd$ \\
    $\Sigma^0$ &&1&$\frac{1}{2}^+$& $1192.642$ & - & $\surd$ \\
    $K^{*+}$ &$K^{*+}(892)$&$\frac{1}{2}$&$1^-$& $891.66$ & $50.8$ & $\surd$ \\
    $K_1$   &$K_1(1270)$ &$\frac{1}{2}$&$1^+$& $1272$ & $90$ & $\surd$ \\
    $N^*_1$&$P_{11}(1440)$ &$\frac{1}{2}$&$\frac{1}{2}^+$& $1440$ & $300$ & - \\
    $N^*_2$&$S_{11}(1535)$ &$\frac{1}{2}$&$\frac{1}{2}^-$& $1535$ & $150$ & - \\
    $N^*_3$&$S_{11}(1650)$&$\frac{1}{2}$&$\frac{1}{2}^-$&$1655$&$165$&$\surd$ \\
    $Y^*_1$ &$S_{01}(1405)$&0&$\frac{1}{2}^-$& $1406.5$ & $50$ & - \\
    $Y^*_2$ &$S_{01}(1670)$&0&$\frac{1}{2}^-$& $1670$ & $35$ & - \\
    $Y^*_3$ &$S_{01}(1800)$&0&$\frac{1}{2}^-$& $1800$ & $300$ & $\surd$ \\
    $Y^*_4$  &$S_{11}(1750)$&1&$\frac{1}{2}^-$& $1750$ & $90$ & - \\
  \end{tabular}
  \end{ruledtabular}
\end{table}

\subsection{Numerical Results}
The particles which might contribute to the kaon photoproduction
near threshold have been listed in Ref. \cite{Cheoun:1996kn}. In 
Table \ref{tab:particle_used} we display this list and indicate
the particle used in the present analysis. It is important to note that
we do not include all those possible resonances because the
inclusion of nucleon resonances below the production threshold
is not possible in our resonance formalism 
(see Sec. \ref{sec:resonance}). Furthermore, from the observation
of the $\chi^2$ obtained from the fits, we learn that the use 
of only one hyperon resonance, i.e. the $S_{01}(1800)$, 
in the background terms is found to be sufficient 
for reproducing the experimental data within their accuracies
(see Subsection \ref{sec:hyperon_effect}). Adding more hyperon resonances
does not significantly improve the $\chi^2$. 

In the present analysis we maintain to use both $K^*(892)$ and 
$K_1(1270)$ meson resonances, since most of previous studies,
e.g. Refs.~\cite{Adelseck:1990ch,Janssen:2001wk}, found that
they are necessary to reproduce the SU(3) values of main coupling
constants in the fits. 
The values of main coupling constants are fixed to the SU(3) 
values, i.e. $g_{K\Lambda N}/\sqrt{4\pi}=-3.80$ and 
$g_{K\Sigma N}/\sqrt{4\pi}=1.20$, as in the case of 
Kaon-Maid \cite{kaon-maid}. The coupling constants of
$K^*$, $K_1$, and the hyperon resonance are fitted to the
available experimental data. In principle, their values can
be estimated from other sources with the help of SU(3) symmetry
and other mechanisms. However, in our analysis, the existence 
of these intermediate states is very important to avoid 
the divergence of the calculated cross section. 
As has been briefly discussed 
in Introduction, in order 
to limit the number of free parameters, we fix the resonance
parameters of $S_{11}(1650)$ to the PDG values given in 
Table \ref{tab:resonance_pdg}. Thus, for the nucleon resonance,
only the phase angle $\phi$ in Eq.~(\ref{eq:em_multipole}) is 
extracted from experimental data. 

\begin{table}[tb]
  \centering
  \caption{Properties of the $S_{11}(1650)$ resonance taken from the 
  Review of Particle Properties \cite{Amsler:2009}.}
  \label{tab:resonance_pdg}
  \begin{ruledtabular}
  \begin{tabular}[c]{ccccccc}
    Resonance & $M_R$ & $\Gamma_R$ & $\beta_K$ & $A_{1/2} (p)$  & Overall & Status\\
     & (MeV) & (MeV) && ($10^{-3}$ GeV$^{-1/2}$)  & status & seen in $K\Lambda$\\
    \hline
    $S_{11}(1650)$ & $1655^{+15}_{-10}$ & $165\pm 20$ &$0.027\pm 0.004$& $+53\pm 16$  & **** & ***\\
  \end{tabular}
  \end{ruledtabular}
\end{table}

Obviously, there is no need to consider hadronic 
form factors here, since contribution from the background 
terms is still controllable. As shown by Refs.~\cite{Adelseck:1990ch,
Hsiao:2000}, the background contribution starts to diverge 
at $W>1900$ MeV. At this stage, even combination of 
several nucleon and hyperon resonances is unable to damp 
the cross section. Thus, the use of hadronic form factors provides 
the only efficient mechanism to overcome this problem. 
Meanwhile, the problem of data inconsistency discussed e.g. in
Refs. \cite{Mart:2006dk,Bydzovsky:2006wy} does not appear in the
energy region of interest, since both CLAS and SAPHIR data agree
each other up to 50 MeV above the threshold.

There are 139 experimental data points in our database for
energies up to 50 MeV above $W_{\rm thr}$, consisting of
115 differential cross sections
\cite{Glander:2003jw,Bradford:2005pt}, 18 recoil 
polarizations \cite{Glander:2003jw,Bradford:2005pt,lleres07}, 
and 6 photon beam asymmetries \cite{lleres07}.
In addition, we have 18 data points of beam-recoil double 
polarizations, $O_x$ and $O_z$, and target asymmetry $T$
from the latest GRAAL measurement \cite{lleres09}. 

A comparison between the extracted coupling constants in the
present analysis and those obtained in previous works is
given in Table \ref{tab:numerical-result}. Our result 
corroborates the finding of previous studies, i.e. experimental
data prefer the PS coupling. The $\chi^2$ of this model is clearly
smaller than that of the PV one, which is an obvious consequence
of the use the SU(3) main coupling constants. In the PV case, 
if we left these coupling constants to be determined by experimental 
data, we would then obtain smaller values like those found in 
previous studies \cite{Hsiao:2000,bennhold:1987}. The phenomenon
originates from the fact that the standard PV Born terms yield 
a larger cross section compared to the standard PS Born terms.
This problem can be alleviated by reducing the absolute values 
of the main coupling constants. In our case, since the coupling
constants are fixed to the SU(3) values, the fit tried to 
reduce this through a destructive interference with
other diagrams, e.g. by increasing the values of the hyperon
resonance $Y_3^*$ coupling constant (see column 3 of Table 
\ref{tab:numerical-result}). Nevertheless, this is insufficient
to compete with the result of the PS model, as shown by the 
larger $\chi^2$ compared to that of the PS model.

\begin{table}[!h]
  \centering
  \caption{The extracted coupling constants of the present work
    (PS, PV, and PS-PV models) compared with those of previous analyses of
    Kaon-Maid \cite{kaon-maid}, 
    Adelseck and Saghai (AS1 and AS2) \cite{Adelseck:1990ch},
    Williams {\it et al} (WJC) \cite{williams}, and Cheoun {\it et al.}
    (CHYC) \cite{Cheoun:1996kn}. Note that the $\chi^2$ of previous
    works are not shown for comparison because they used different
    experimental database.}
  \label{tab:numerical-result}
  \begin{ruledtabular}
  \begin{tabular}[c]{lrrrcrrrc}
    Coupling Constants & PS~ & PV~ & PS-PV & Kaon-Maid& AS1 & AS2 & WJC & CHYC \\
\hline
  $g_{K \Lambda N}/\sqrt{4\pi}$&$-3.80$ &$-3.80$ &$-3.80$ &$-3.80$ & $ -4.17$ & $-4.26 $ & $-2.38$ & varies\\
  $g_{K \Sigma N}/\sqrt{4\pi}$ &$ 1.20$ &$ 1.20$ &$ 1.20$ &$~~1.20$ & $1.18$ & $1.20$ &0.23 & varies\\
  $G^{V}_{K^{*}}/4\pi$ &$-0.65$ &$-0.79$ &$-0.65$&$-0.79$& $-0.43$ & $-0.38$ &$-0.16$ & $-0.09$ \\
  $G^{T}_{K^{*}}/4\pi$ &$ 0.29$ &$-0.04$ &$ 0.28$&$-2.63$& $0.20$ & $0.30$ &0.08& $-0.17\sim -0.36$ \\
  $G^{V}_{K_1}/4\pi$   &$0.42$  &$ 1.19$ &$ 0.42$&$~~3.81$& $-0.10$ & $-0.06$ &0.02 &$-0.06$\\
  $G^{T}_{K_1}/4\pi$   &$-3.17$ &$-0.68$ &$ -3.16$&$-2.41$& $-1.21$ & $-1.35$ &0.17 &$-0.11\sim -0.23$\\
  $G_{Y_1^*}/\sqrt{4\pi}$ &-&-&-&-& - & $-2.47$ & $-0.10$ &-\\
  $G_{Y_2^*}/\sqrt{4\pi}$ &-&-&-&-&$-3.17$ & - & - &- \\
  $G_{Y_3^*}/\sqrt{4\pi}$ &$-4.93$&$-10.00$&$-5.93$&-& - & - & - &-\\
  $\phi$ (deg) & 218 & 202 &218& -& - & - & - & -\\
  $\lambda_{\rm PV}$ & - & - & 0.13 & - & - & - & -& -\\
  \hline
  $\chi^2$   & 127.9 & 212.2 &127.8& & &&&\\
  $\chi^2/N$ & 0.920 & 1.526 &0.920& & &&&\\
  \end{tabular}
  \end{ruledtabular}
\end{table}

The model with a mixed PS-PV coupling is obtained if we consider 
the mixing parameter $\lambda_{\rm PV}$ in Eqs. (\ref{eq:born_pv1}) -
(\ref{eq:born_pv6}) as a free parameter in the
fitting process. In this case the result is shown in the fourth column
of Table III. It is obvious that this PS-PV (or hybrid) 
model is practically identical
to the PS model, since all coupling constants and the value of $\chi^2/N$ 
are the same as in the PS model. Therefore, in the following discussion
we will only focus on the results of PS and PV models. 

Although we used different set of resonances, the extracted coupling 
constants of our PS model seem to be closer to those of AS1 
and AS2 models, 
indicating the consistency of our work. The small differences between
our, AS1, and AS2 coupling constants might originate from the different
experimental data and resonance configuration used. Especially remarkable
is the value of the hyperon resonance $Y^*_3$ coupling constant which is
closer to the $Y^*_2$  and $Y_1^*$ coupling constants of the 
AS1 and AS2 model, respectively, but differs by more than one order
of magnitude to that of the WJC coupling ($Y_1^*$). 

\begin{figure}[!h]
  \begin{center}
    \leavevmode
    \epsfig{figure=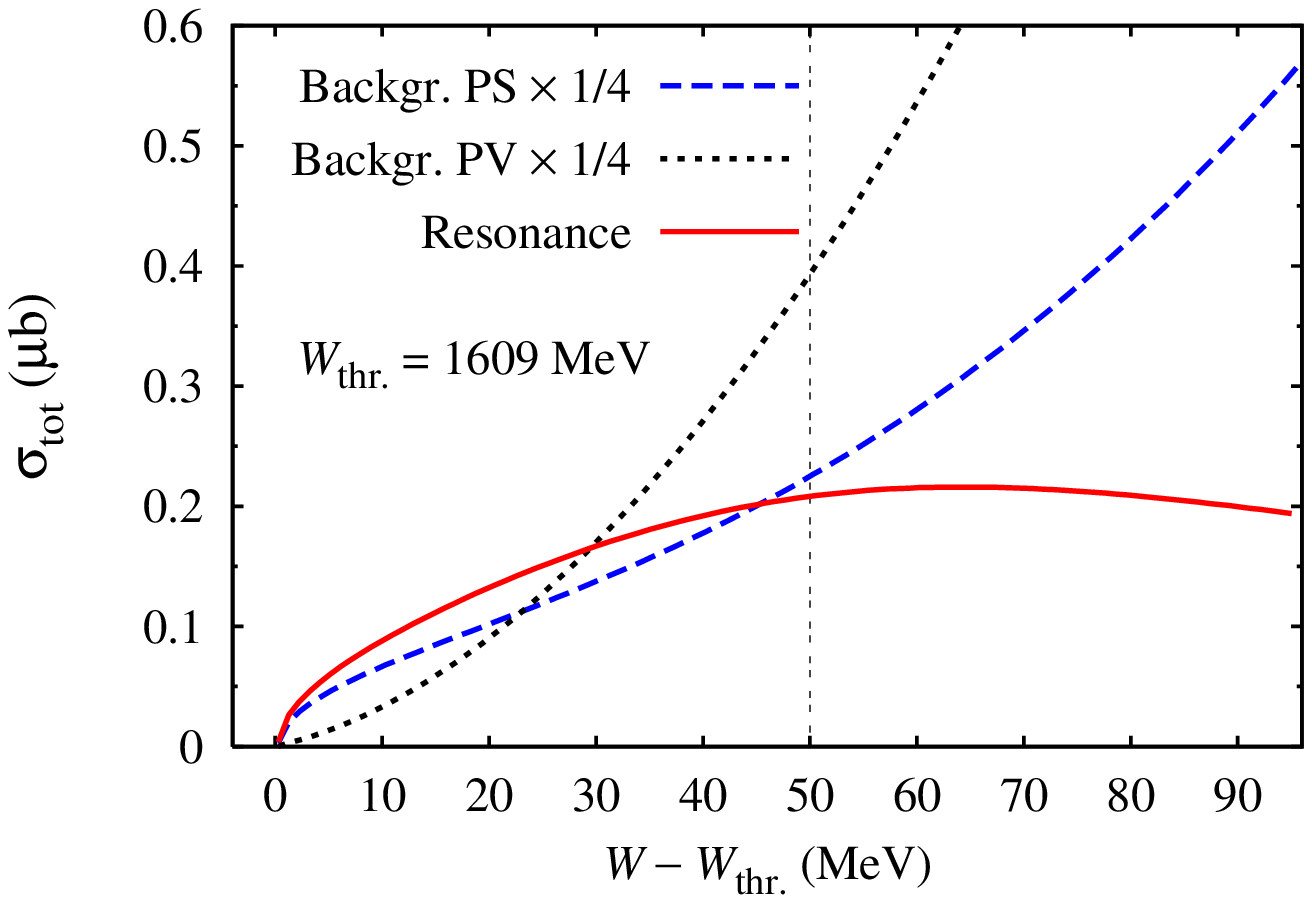,width=100mm}
    \caption{(Color online) Contribution of the background terms and the 
      $S_{11}(1650)$ resonance amplitudes to the total cross section 
      for the PS and PV couplings. Note that contribution of 
      the $S_{11}(1650)$ resonance is the same in both couplings 
      since the resonance parameters are fixed to the PDG values
      \cite{Amsler:2009}. The vertical line at 50 MeV above the
      production threshold indicates the upper limit of
      the energy of the present analysis. }
   \label{fig:contrib} 
  \end{center}
\end{figure}

\begin{figure}[!h]
  \begin{center}
    \leavevmode
    \epsfig{figure=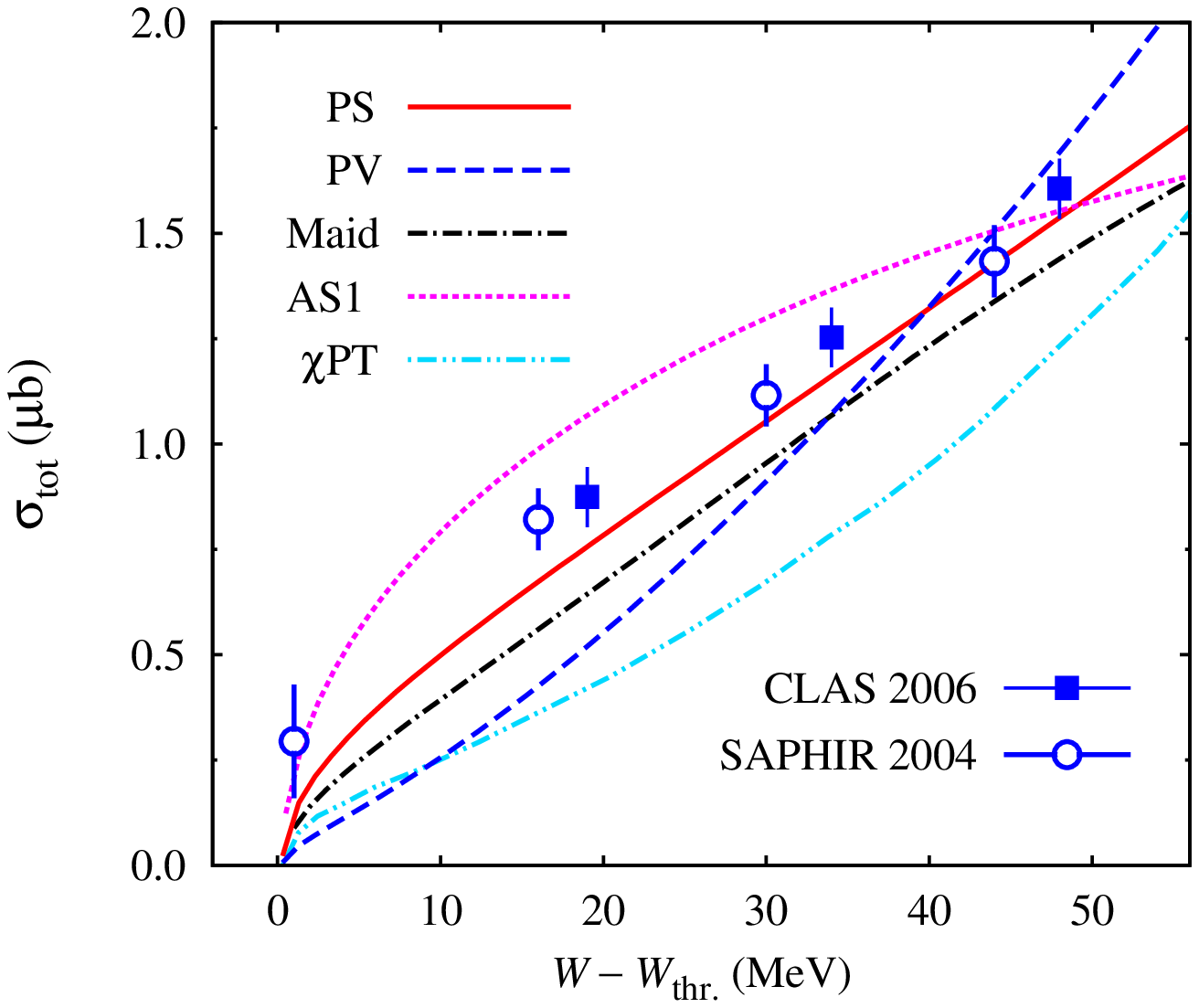,width=90mm}
    \caption{(Color online) Comparison between total cross sections 
      calculated from the PS, PV, AS1 \cite{Adelseck:1990ch}, 
      and Kaon-Maid \cite{kaon-maid} models with the result 
      of the chiral perturbation theory ($\chi$PT) 
      \cite{Steininger:1996xw} and 
      available experimental data from the SAPHIR \cite{Glander:2003jw}
      and  CLAS \cite{Bradford:2005pt} collaborations. 
      Note that all data shown in this figure
      were not used in the fitting process of the PS, PV, and Kaon-Maid models.
      }
   \label{fig:total} 
  \end{center}
\end{figure}

To investigate the role of the nucleon resonance $S_{11}(1650)$
in kaon photoproduction near threshold we compare contribution 
of this resonance with contributions of the backgrounds of the 
PS and PV models to the total cross section in Fig. \ref{fig:contrib}. 
It is found that the $S_{11}(1650)$ contributes more than 20\% 
to the total cross section in the PS model. In the PV model
its contribution varies from about 15\%  up to 70\%, depending
on the energy. In 
Ref. \cite{Mart:2006dk} it is shown that this resonance is only
significant in the model that fits to the SAPHIR data, in which
around 20\% of the total cross section comes from the contribution
of this resonance. In the model that fits to the CLAS data, the
effect of this resonance is negligible (see Fig. 13 of 
Ref. \cite{Mart:2006dk}). Note that the background terms of the
multipole model presented in Ref. \cite{Mart:2006dk}) is based
on the PS coupling. 
We also note that the background contribution
of PV model starts to dramatically increase at $W=25$ MeV above the
threshold, as shown in Fig. \ref{fig:contrib}.
This indicates the deficiency of the PV model at higher energies,
as clearly shown in Fig. \ref{fig:total}. In previous analyses
it was found that the deficiency of the PV model is caused
by the fact that its background terms are found to be too large
compared to those of the PS model 
\cite{bennhold:1987,Cheoun:1996kn,thom}. 
Our result is therefore consistent
with previous analyses. Furthermore, from Fig. \ref{fig:contrib}
we found that between threshold and $W=25$ MeV the PV background terms
surprisingly 
yield smaller contribution to the cross section. Thus, very close
to threshold the use of SU(3) main coupling constants should
be also acceptable in the PV model.

\subsection{Comparison with Experimental Data}

\begin{figure}[!h]
  \begin{center}
    \leavevmode
    \epsfig{figure=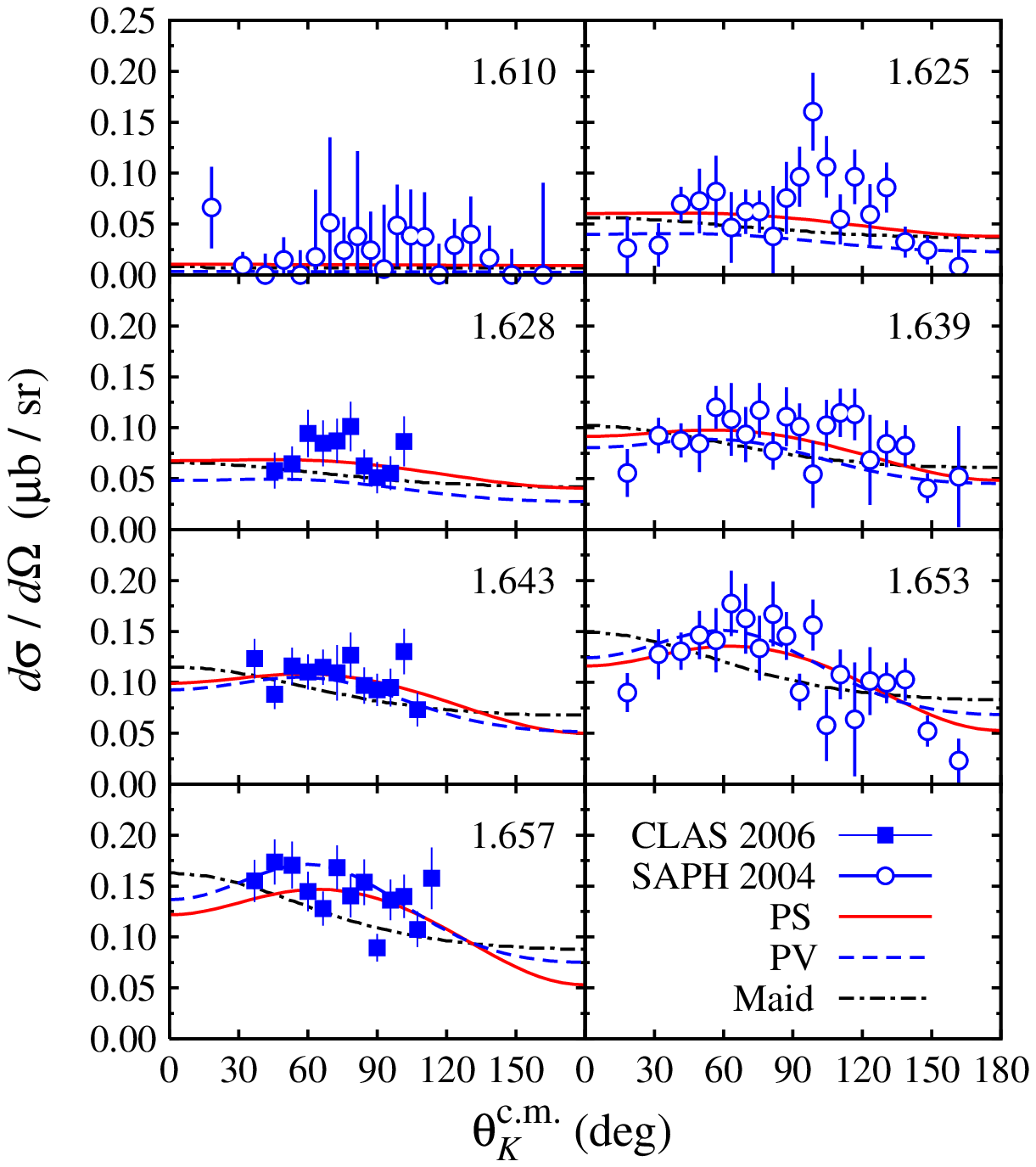,width=100mm}
    \caption{(Color online) Comparison between angular distributions of 
             differential cross section 
             calculated from the PS, PV, and Kaon-Maid 
             \cite{kaon-maid} models with experimental data 
             from the SAPHIR \cite{Glander:2003jw} and 
             CLAS \cite{Bradford:2005pt} collaborations.
             The corresponding total c.m. energy $W$ (in GeV) is
             shown in each panel. Experimental data displayed 
             in this figure were used
             in the fits of the PS and PV models.}
   \label{fig:dif_th} 
  \end{center}
\end{figure}

In Fig. \ref{fig:total} we compare total cross sections 
predicted by the PS, PV, AS1, and 
Kaon-Maid \cite{kaon-maid} models with experimental data and 
prediction of the chiral perturbation theory \cite{Steininger:1996xw}. 
Obviously, the PS model is the best model for kaon photoproduction
near threshold. The PV result underestimates the data up to
$W=40$ MeV above threshold. Kaon-Maid underpredicts the data
by about 20\% in the whole energy region of interest. The similar
phenomenon is also shown by the prediction of chiral perturbation 
theory, i.e., it also underpredicts experimental data by 
about 30\% - 50\%. Interestingly, we observe that the prediction
of the PV model is very close to the prediction of the chiral perturbation 
theory up to $W=10$ MeV above threshold. This fact originates from
the small background terms of the PV model in this energy region, 
as discussed in the
previous subsection. The AS1 model seems to overestimate most of
the experimental data shown. Note that we did not use these total
cross section data in our fits, because they are less accurate
than the differential ones. Especially in the case of the CLAS
data, where the angular distribution coverage is more limited
than in the case of SAPHIR data. 

\begin{figure}[!h]
  \begin{center}
    \leavevmode
    \epsfig{figure=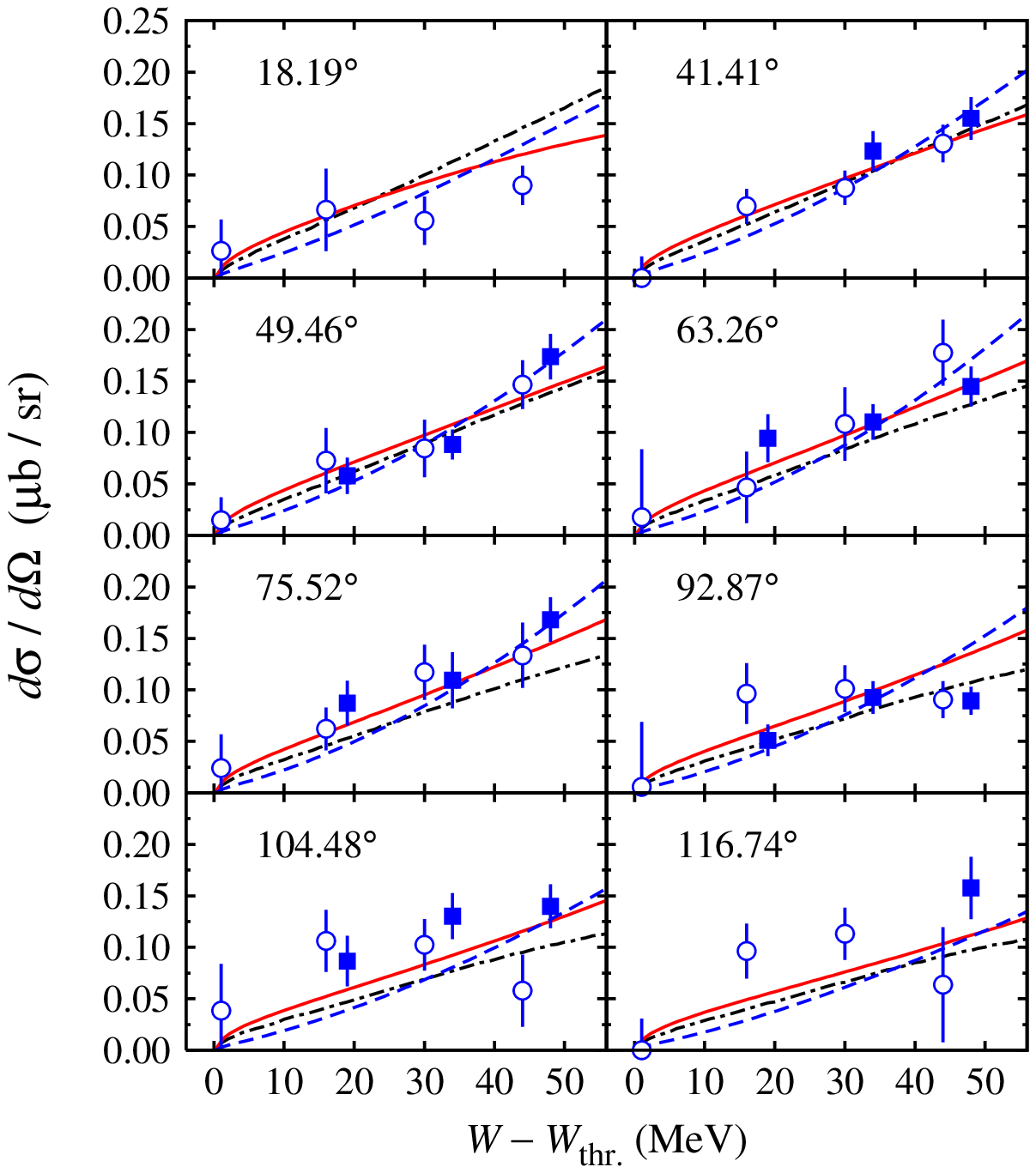,width=100mm}
    \caption{(Color online) As in Fig.~\ref{fig:dif_th}, 
             but for the total c.m. energy
             distribution. The corresponding kaon scattering 
             angle $\theta_K^{\rm c.m.}$ is shown in each panel. }
   \label{fig:dif_w} 
  \end{center}
\end{figure}

The angular distribution of differential cross section shows 
a certain structure (see Fig. \ref{fig:dif_th}). This structure
is more obvious in the case of SAPHIR data. This structure seems
to be missing in the Kaon-Maid, as clearly seen at $W=1.653$ MeV and
1.657 MeV. We also note that both AS1 and AS2 cannot produce
this structure. Nevertheless, due to 
their limited accuracies, present experimental data do not allow
for further analysis of this structure. Even the difference between
the PS and PV models cannot be resolved at present.
Therefore, for investigation of kaon photoproduction
near threshold, experimental measurement of differential cross
section with an accuracy of about 5\% in the whole angular distribution
is recommended. Especially important is  the production 
at threshold, where isobar models seem to produce constant and 
structureless differential cross sections.

The energy distribution of differential cross section is shown in
Fig. \ref{fig:dif_w}. Here we see that the agreement between model
calculations and experimental data is quite satisfactory. 
The small deviations at $\theta_K^{\rm c.m.}=104.48^\circ$ and $116.74^\circ$
originate from the influence of backward angles data, specifically
from SAPHIR data (see Fig. \ref{fig:dif_th}), which decrease the
cross section at this kinematics during the fitting process.

\begin{figure}[!h]
  \begin{center}
    \leavevmode
    \epsfig{figure=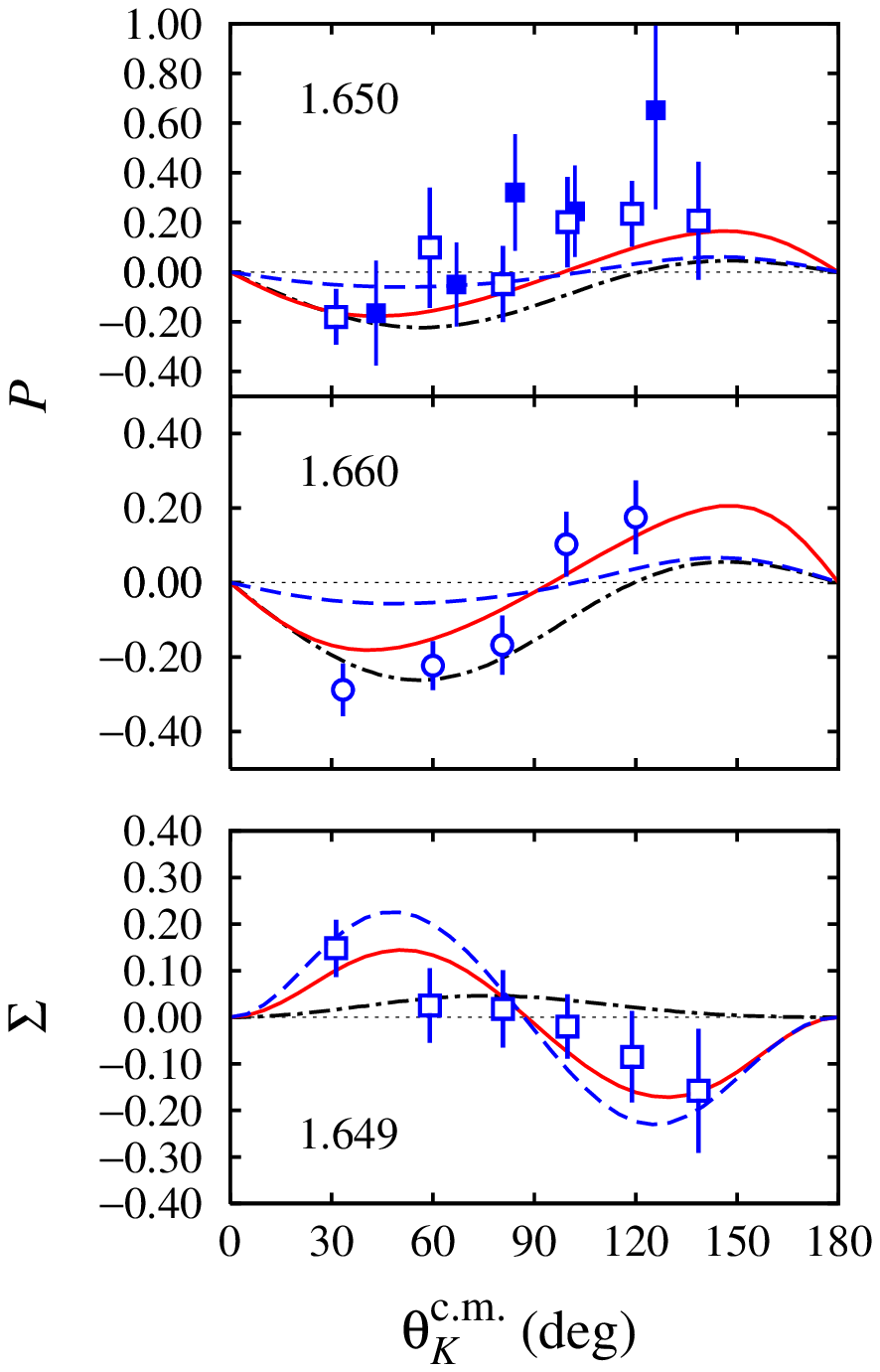,width=70mm}
    \caption{(Color online) Recoil polarization $P$ and photon asymmetry 
             $\Sigma$ calculated from the PS, 
             PV, and Kaon-Maid models \cite{kaon-maid} 
             compared with experimental data 
             from the SAPHIR \cite{Glander:2003jw} (open circles),
             CLAS \cite{Bradford:2005pt} (solid squares), and 
             GRAAL \cite{lleres07} (open squares) collaborations.
             Notation for the curves is as in Fig. \ref{fig:dif_th}.
             The corresponding total c.m. energy $W$ (in GeV) is
             shown in each panel.}
   \label{fig:polar} 
  \end{center}
\end{figure}

The results for recoil polarization and photon asymmetry observables
are shown in Fig. \ref{fig:polar}. It is widely known that these
observables are quite sensitive to the choice of the resonance 
configuration. Again, we see that the agreement of the PS model
is better than both PV and Kaon-Maid models.

\begin{figure}[!h]
  \begin{center}
    \leavevmode
    \epsfig{figure=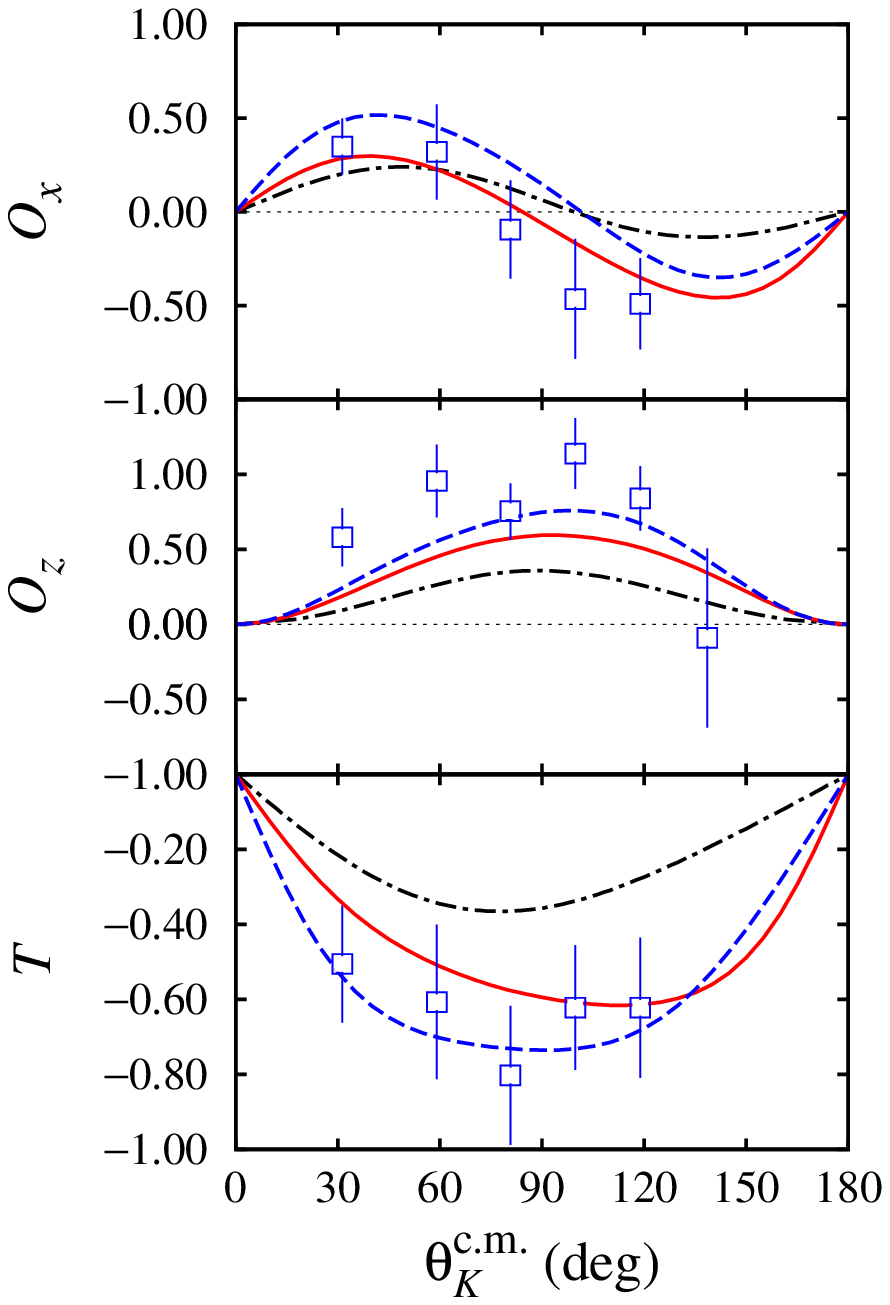,width=70mm}
    \caption{(Color online) The beam-recoil double polarization observables $O_x$
             and $O_z$, along with target asymmetry $T$ predicted
             by the PS, PV, and Kaon-Maid models \cite{kaon-maid} 
             compared with experimental data 
             from the GRAAL \cite{lleres09} collaboration.
             Notation of the curves is as in Fig. \ref{fig:dif_th}.
             Note that all experimental data shown in this figure
             were not used in the fitting process of all models.}
   \label{fig:ox_oz} 
  \end{center}
\end{figure}

\subsection{The Beam-Recoil and Target Polarizations}

The new experimental data for the beam-recoil and target 
polarizations have been recently released by the GRAAL
collaboration \cite{lleres09}. These data require special
attention because their definition of coordinate systems
is slightly different from ours. The GRAAL data are obtained
by using the coordinate system used by Adelseck and Saghai 
\cite{Adelseck:1990ch}. In the present work we used the coordinate
system of Kn\"ochlein {\it et al.} \cite{germar}. We note 
that the difference between the two coordinate systems 
leads to different signs of both $O_x$ and
$O_z$ observables, but the same sign of the target asymmetry
$T$. At this stage, it is important to remind the reader that
already in the ''classical'' paper of Barker, Donnachie and 
Storrow in 1975 \cite{Barker:1975bp} a clear definition of
signs and coordinate systems for this purpose has been given.
In the work of Kn\"ochlein {\it et al} \cite{germar} these
coordinate systems have been adopted. 
However, in the recent work of Sandorfi {\it et al.}
\cite{Sandorfi:2009xe} new expressions that allow
a direct calculation of matrix elements with arbitrary
spin projections are presented and used to clarify sign
differences that exist in the literature. By comparing
with Kaon-Maid (as well as SAID), it is found 
that within this convention the implied definition of six 
double-polarization observables, i.e. $H$, $L_{x'}$, 
$C_{x'}$, $C_{z'}$, $O_{x'}$, and $O_{z'}$, are the negative of
what has been used in comparing to recent experimental data.
To comply with this finding we do not flip the original sign 
of the experimental data of the GRAAL collaboration \cite{lleres09}
as shown in Fig. \ref{fig:ox_oz}.
Note that the data shown in Fig. \ref{fig:ox_oz} 
are not used in our analysis. Therefore, the calculated
polarization observables shown here are pure prediction.
Surprisingly, the predicted $O_{x}$ and $O_{z}$, as well as
the target asymmetry $T$, are in perfect agreement 
with experimental data. This is true for 
both PS and PV models, although the presently available data
cannot resolve the difference between the two models. 
We observe that the present calculation is still 
consistent with the result of Kaon-Maid model \cite{kaon-maid}.
We have also compared our result for the $O_x$ observable with 
that obtained by 
Adelseck and Saghai (see Fig. 11 of Ref. \cite{Adelseck:1990ch})
and found that our result is also consistent, especially at the
backward direction, where the sign of $O_x$ becomes negative.
To conclude this subsection, we would like to say that our
result corroborates the finding of Sandorfi {\it et al.}
\cite{Sandorfi:2009xe}.
\subsection{Effects of the Hyperon Resonance}
\label{sec:hyperon_effect}
The importance of the hyperon resonance in the background amplitude
of isobar model for kaon photoproduction was pointed out by 
Adelseck and Saghai \cite{Adelseck:1990ch}. They used the 
$S_{01}(1670)$ resonance in the AS1 model and the $S_{01}(1405)$ 
in the AS2 model (see Table \ref{tab:numerical-result}). 
By including $S_{01}(1405)$ or 
$S_{01}(1670)$ they were able to fit the hitherto available
experimental data and simultaneously reproduce the SU(3) values
of main coupling constants $g_{K\Lambda N}$ and $g_{K\Sigma N}$. 
Later, Janssen {\it et al.} 
\cite{Janssen:2001wk} found that the use of hyperon resonances 
$S_{01} (1800)$ and $P_{01}(1810)$, simultaneously, is 
desired in order to obtain the hadronic form factor 
cut-off greater than 1500 MeV in their model. The larger value of 
cut-off is desired in the kaon photoproduction because they argued 
that the value of $800$ MeV found in Ref. \cite{missing-d13} 
is rather unrealistic. In contrast to this, Kaon-Maid does not
use any hyperon resonance in its background terms.

\begin{table}[!t]
  \centering
  \caption{Effects of the hyperon resonance 
    inclusion on the  total $\chi^2$ 
    obtained from fits to the photoproduction data.}
  \label{tab:chi2_hyp_effect}
  \begin{ruledtabular}
  \begin{tabular}[c]{cccccc}
    Resonance&-&$S_{01}(1405)$&$S_{01}(1670)$&$S_{01}(1800)$&$S_{11}(1750)$\\
    \hline
    $\chi^2$ & 189.4 & 136.7 & 135.7 & 127.9 & 133.7\\
  \end{tabular}
  \end{ruledtabular}
\end{table}

\begin{figure}[!t]
  \begin{center}
    \leavevmode
    \epsfig{figure=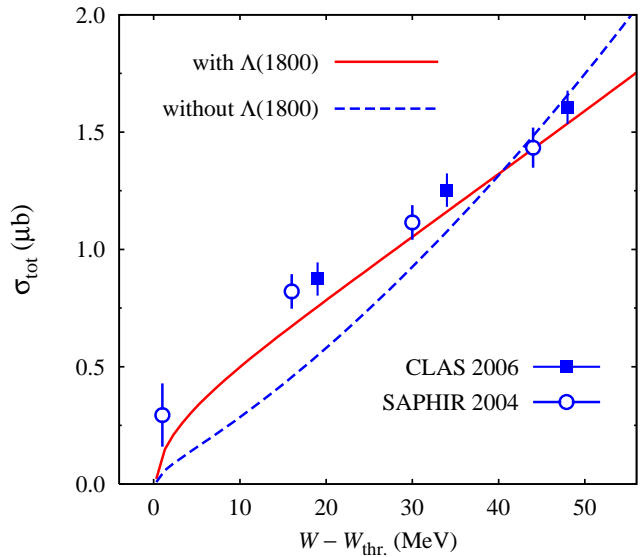,width=90mm}
    \caption{(Color online) Effect of the hyperon resonance 
      $\Lambda(1800)$ inclusion on the
      total cross sections in the PS model.}
   \label{fig:effect_total} 
  \end{center}
\end{figure}

In the present work we have tested the sensitivity
of our fits to all four 
hyperon resonances listed in Table \ref{tab:particle_used}. 
The values of $\chi^2$ obtained from these fits are
listed in Table \ref{tab:chi2_hyp_effect}. It is obvious that
significant improvement in the $\chi^2$ would be obtained once we included 
this resonance, especially when the $S_{01} (1800)$ is used.
This result corroborates the finding of Janssen {\it et al.} 
\cite{Janssen:2001wk}, since the large hadronic cut-off 
requires another mechanism to damp the cross section 
at high energies. This is achieved by a destructive 
interference between the hyperon resonance contribution
and other background terms. 
Such an interference is also observed in the present
work, as obviously indicated by the total cross sections
in Fig. \ref{fig:effect_total}. Although underpredicts 
experimental data at $W-W_{\rm thr}< 40$ MeV, the model
that excludes the $S_{01} (1800)$ resonance starts to 
overpredict the data, and therefore starts to diverge, 
at  $W-W_{\rm thr}> 40$ MeV.

\begin{figure}[!t]
  \begin{center}
    \leavevmode
    \epsfig{figure=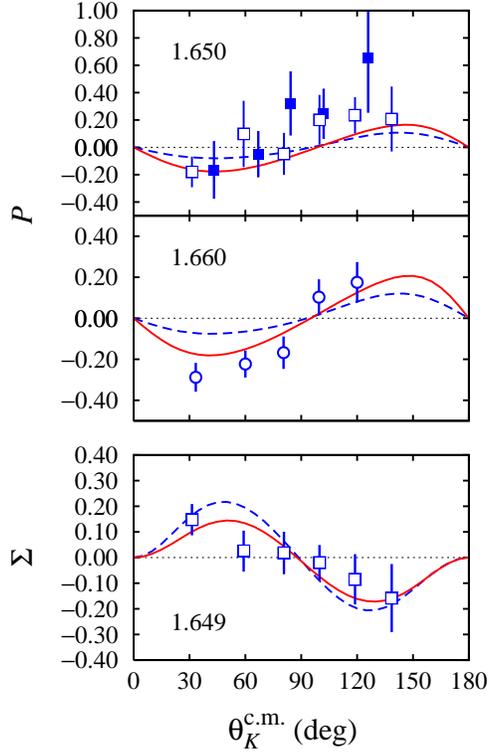,width=70mm}
    \caption{(Color online) Effect of the hyperon resonance 
      $\Lambda(1800)$ inclusion on the
      recoil polarization and target asymmetry 
      in the PS model. Notation of the curves is 
      as in Fig. \ref{fig:effect_total}. Experimental
      data are as in Fig. \ref{fig:polar}.}
   \label{fig:effect_polar} 
  \end{center}
\end{figure}

The improvement obtained by including this resonance 
is not only observed in the total as well as differential
cross sections, but also found in the calculated recoil
polarizations and the target asymmetry as shown in 
Fig. \ref{fig:effect_polar}. It is obvious from
this figure that including this hyperon resonance 
results in a perfect agreement between our PS model
and experimental data, especially the SAPHIR and
GRAAL ones.

\section{Constraint from Kaon Electroproduction}
\label{sec:electroproduction}
\begin{figure}[!t]
  \begin{center}
    \leavevmode
    \epsfig{figure=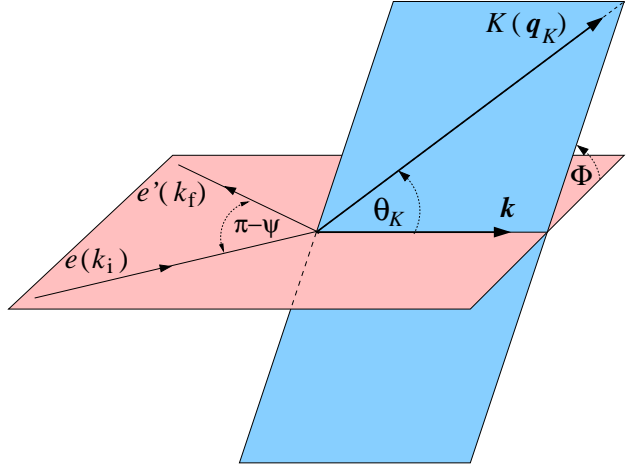,width=90mm}
    \caption{(Color online) Kinematic variables for kaon electroproduction 
      on nucleons.}
   \label{fig:kin-var-electro}
  \end{center}
\end{figure}

For a polarized electron beam with helicity $h$ and no target
or recoil polarization, the cross section of kaon electroproduction 
on a nucleon can be written as
\begin{eqnarray}
  \label{eq:virt-cross}
  \frac{d\sigma_v}{d\Omega_K} &=& \frac{d\sigma_{\rm T}}{d\Omega_K}+
  \epsilon\,\frac{d\sigma_{\rm L}}{d\Omega_K}+\epsilon\,
  \frac{d\sigma_{\rm TT}}{d\Omega_K}\cos 2\Phi 
  + \sqrt{\epsilon
    (1+\epsilon)}\,\frac{d\sigma_{\rm LT}}{d\Omega_K}\,\cos\Phi 
  \nonumber\\&&
  + h\sqrt{\epsilon
    (1-\epsilon)}\,\frac{d\sigma_{\rm LT'}}{d\Omega_K}\,\sin\Phi ~,
\end{eqnarray}
where $\Phi$ is the angle between the electron (scattering) and 
hadron reaction planes, and $\epsilon$ is the transverse polarization
of the virtual photon
\begin{eqnarray}
  \label{eq:polar-virt-photon}
  \epsilon = \biggl(1 - \frac{2{\bf k}^{2}}{k^{2}} 
\tan^{2}(\psi /2) \biggr)^{-1}~ ,
\end{eqnarray}
with $\psi$ the electron scattering angle. All kinematical variables
given in Eqs.~(\ref{eq:virt-cross}) and (\ref{eq:polar-virt-photon})
are shown in Fig.~\ref{fig:kin-var-electro}.

The subscripts T, L, and TT in Eq.~(\ref{eq:virt-cross}) 
stand for transversely unpolarized, longitudinally polarized, 
and transversely polarized virtual photons. The subscripts 
LT and LT' refer to the interference between transversely and 
longitudinally polarized virtual photons. Although they
are generated by the same interference between longitudinal 
and transverse components of the hadronic and leptonic currents,
they are different because they are generated by the real 
and imaginary parts of this interference term, respectively. 
The individual terms may be expressed 
in terms of the functions $A_i$ calculated in
Sections \ref{sec:ps}--\ref{sec:pv}.

In the following discussion we adopt the notation of Refs. 
\cite{Ambrozewicz:2006zj,Nasseripour}, i.e.
\begin{subequations}
\label{eq:notation-cs}
\begin{eqnarray}
  \sigma_{\rm T} &=& \frac{d\sigma_{\rm T}}{d\Omega_K}\, ,\\
  \sigma_{\rm L} &=& \frac{d\sigma_{\rm L}}{d\Omega_K} \, ,\\
  \sigma_{\rm U} &=& \sigma_{\rm T} + \epsilon\,\sigma_{\rm L} \, ,\\
  \sigma_{\rm TT} &=& \frac{d\sigma_{\rm TT}}{d\Omega_K}\, ,\\
  \sigma_{\rm LT} &=& \frac{d\sigma_{\rm LT}}{d\Omega_K} \, ,\\
  \sigma_{\rm LT'} &=& \frac{d\sigma_{\rm LT'}}{d\Omega_K} \, ,
\end{eqnarray}
\end{subequations}
where $\sigma_{\rm U}$ refers to the unpolarized 
differential cross section.

The important ingredients in the electroproduction process
are the electromagnetic form factors. Traditionally, to 
extend a photoproduction model to the case of virtual photon
we may use the same hadronic coupling constants 
extracted from photoproduction data and put the appropriate 
electromagnetic form factors in the electromagnetic vertices.
Since we are not in the position to investigate the effect
of electromagnetic form factors in the kaon electroproduction
process we use the standard Dirac and Pauli form factors for 
the proton, expressed 
in terms of the Sachs form factors, 
\begin{subequations}
\label{eq:F1_F2}
\begin{eqnarray}
F_{1}^p(Q^{2}) & = & \biggl[ G_{E}(Q^{2}) + \frac{Q^{2}}{4 m_{p}^{2}}
G_{M}(Q^{2}) \biggr] \biggl( 1 + \frac{Q^{2}}{4 m_{p}^{2}} \biggr)^{-1}~ ,\\
F_{2}^p(Q^{2}) & = & \biggl[ G_{M}(Q^{2}) - 
G_{E}(Q^{2}) \biggr] \biggl[ \kappa_p \biggl( 1 + \frac{Q^{2}}{4 
m_{p}^{2}} \biggr) \biggr]^{-1}~ ,
\end{eqnarray}
\end{subequations}
where $\kappa_p$ is the anomalous magnetic moment of the proton
and $Q^2=-k^2$. In the momentum transfer region of interest, 
the Sachs form factors of the proton $G_{E}^{p}(Q^{2})$ and 
$G_{M}^{p}(Q^{2})$ can be described by the standard dipole form 
\begin{eqnarray}
  \label{eq:dipole_ff}
  G_{E}^{p}(Q^{2})& =& \frac{G_{M}^{p}(Q^{2})}{1 + \kappa_{p}} ~=~
  \left(1+\frac{Q^{2}}{0.71~{\rm GeV}^{2}} \right)^{-2}  ~.
\end{eqnarray}
The kaon form factor is taken from the work of Cardarelli {\it et al.}
\cite{cardarelli}, i.e.
\begin{eqnarray}
  F^{K^+}(Q^2) = \frac{a}{1+Q^2/\Lambda_1^2}+\frac{1-a}{(1+Q^2/\Lambda_2^2)^2} ,
\end{eqnarray}
with $a=0.398$, $\Lambda_1= 0.642$ GeV, and $\Lambda_2= 1.386$ GeV.
We note that this form factor has been used in the calculation
of kaon electroproduction in Ref.~\cite{david}.
For kaon resonances $K^*$ and $K_1$ we use the standard monopole form 
factor
\begin{eqnarray}
  F^{K^*}(Q^2) = \left(1+\frac{Q^2}{\Lambda_{K^*}^2}\right)^{-1} ~~~~~,~~~~~
  F^{K_1}(Q^2) = \left(1+\frac{Q^2}{\Lambda_{K_1}^2}\right)^{-1} ~,
\label{eq:monopole}
\end{eqnarray}
where the cutoffs $\Lambda_{K^*}$ and $\Lambda_{K_1}$ are 
considered as free parameters. On the other hand for the
form factors of hyperon and hyperon resonances we 
adopt the standard dipole form
\begin{eqnarray}
  F^{Y}(Q^2) = \left(1+\frac{Q^{2}}{\Lambda_Y^{2}} \right)^{-2} ~~~~,~~~~
  F^{Y^*}(Q^2) = \left(1+\frac{Q^{2}}{\Lambda_{Y^*}^2} \right)^{-2} ~,
\end{eqnarray}
where $\Lambda_{Y}$ and $\Lambda_{Y^*}$ are fitted to experimental data.
In the case of PV coupling we also assume a monopole form as 
in Eq.~(\ref{eq:monopole}) for the electromagnetic form factor
in the contact diagram vertex. 
Finally, the dependence of the electric 
and scalar multipoles on $Q^2$ is taken as~\cite{hanstein99}
\begin{subequations}
\begin{eqnarray}
  E_{0+}(W,Q^2) &=& E_{0+}(W)(1+\alpha Q^2)\exp(-\beta Q^2) ~,\\
  S_{0+}(W,Q^2) &=& S_{0+}(W)(1+\alpha Q^2)\exp(-\beta Q^2) ~,
\end{eqnarray}
\end{subequations}
where $\alpha$ and $\beta$ are extracted from the fitting process.

We note that there are no kaon electroproduction data
available very close to threshold. The lowest energy where the latest
experiment has been performed at JLab is $W=1.65$ GeV. In fact, 
this value comes from the bin center of the collected data with 
$W$ spans from 1.60 GeV to 1.70 GeV \cite{Ambrozewicz:2006zj}. 
The extracted 
data are $\sigma_{\rm U}$, $\sigma_{\rm TT}$, $\sigma_{\rm LT}$, which in total
consist of 108 data points at $W=1.65$ GeV and $Q^2=0.65 - 2.55$
GeV$^2$. In addition, there have been also data available for the 
polarized structure function $\sigma_{\rm LT'}$. To this end 
there are 12 extracted points at the same $W$ with $Q^2=0.65$ GeV$^2$
and $1.00$ GeV$^2$ \cite{Nasseripour}. 

Surprisingly, fitting the 120 electroproduction data points by 
adjusting 7 longitudinal parameters given above, 
i.e., $S^0_{1/2+}(p)$, $\alpha$,
$\beta$, $\Lambda_Y$, $\Lambda_{Y_3^*}$, $\Lambda_{K^*}$ and 
$\Lambda_{K_1}$, is almost impossible, since in this case
we obtain $\chi^2/N\approx 4000$. We have checked our Fortran
code and found that this is caused by the large contribution
from the background as the momentum transfer $Q^2$ increases, 
especially from the $K^*$ and $K_1$ 
intermediate states. There is of course many possible ways to limit
their contributions. One of them is by introducing
the same form factor as the one we have used 
in the multipoles. Thus, for instance, the $K^*$ 
intermediate state has 
\begin{eqnarray}
  \label{eq:ff_exp}
  F^{K^*}(Q^2) &=& (1+\alpha_{K^*} Q^2)\exp(-\beta_{K^*} Q^2)~. 
\end{eqnarray}
The result is shown in Table \ref{tab:numerical-result2},
where we have also used this form factor [Eq. (\ref{eq:ff_exp})] 
in the ${Y_3^*}$ intermediate state, in order to obtain the
lowest $\chi^2$. 

\begin{table}[!h]
  \centering
  \caption{Extracted parameters from fit to electroproduction data (PS)
    and to both electroproduction and photoproduction data bases (PS1).
    The hadronic coupling constants in the PS model
    (marked with $\dagger$) were obtained from fit to photoproduction
    data and are fixed during the fit to electroproduction data.
    All fits use the PS coupling and the SU(3) predictions
    of the main coupling constants,
    $g_{K \Lambda N}/\sqrt{4\pi}=-3.80$ and $g_{K \Sigma N}/\sqrt{4\pi}= 1.20$.
    }
  \label{tab:numerical-result2}
  \begin{ruledtabular}
  \begin{tabular}[c]{lcc}
    Coupling Constants & PS & PS1  \\
\hline
  $G^{V}_{K^{*}}/4\pi$&$-0.65^\dagger~$&$-0.13~~$ \\
  $G^{T}_{K^{*}}/4\pi$&$0.29^\dagger$&$-0.07~~$ \\
  $G^{V}_{K_1}/4\pi$  &$0.42^\dagger$&$ 0.06$ \\
  $G^{T}_{K_1}/4\pi$  &$-3.17^\dagger~$&$-0.04~~$ \\
  $G_{Y_3^*}/\sqrt{4\pi}$ &$-4.93^\dagger~$&$-5.00~~$\\
  $\phi$ (deg) & $218^\dagger$ & 322 \\
  $S_{1/2+}(p)$ ($10^{-3}$ GeV$^{-1/2}$) &21.4 & $-29.0~~$\\
  $\alpha$ (GeV$^{-2}$)&4.62& $ 3.10$\\
  $\beta$ (GeV$^{-2}$) &1.13& $ 0.97$\\
  $\Lambda_{Y}$ (GeV) &1.10& $ 0.85$\\
  $\alpha_{Y_3^*}$ (GeV$^{-2}$)&0.76& -\\
  $\beta_{Y_3^*}$ (GeV$^{-2}$) &1.75& -\\
  $\alpha_{K^*}$ (GeV$^{-2}$)&0.00& -\\
  $\beta_{K^*}$ (GeV$^{-2}$) &5.50& -\\
  $\alpha_{K_1}$ (GeV$^{-2}$)&0.00& -\\
  $\beta_{K_1}$ (GeV$^{-2}$) &5.50& -\\
  $\Lambda_{Y_3^*}$ (GeV)&-&$ 0.73$\\
  $\Lambda_{K^*}$ (GeV) &-& $ 1.74$\\
  $\Lambda_{K_1}$ (GeV) &-& $ 1.99$\\
  \hline
  $\chi^2/N$ &1.27 &1.22 \\
  \end{tabular}
  \end{ruledtabular}
\end{table}

As shown in the second column of Table \ref{tab:numerical-result2}, 
fitting the 120 electroproduction data points by 
adjusting 7 longitudinal parameters leads to
non-conventional electromagnetic form factors 
of $K^*$ and $K_1$, because they practically have 
the form of $e^{-5.5Q^2}$. Note that the choice of the 
upper limit 5.5 is clearly trivial. However, it would not
substantially change the conclusion if we increased
it, since in the fitting process this value will always 
increase to the upper limit with only small effect on 
the $\chi^2$. Therefore, the use of the exponential
type of form factor given by Eq.~(\ref{eq:ff_exp})
will in principle remove contributions of the 
$K^*$ and $K_1$ intermediate states at finite $Q^2$.

If we compare the extracted $K^*$ and $K_1$ coupling 
constants from fit to photoproduction data (second
columns of Tables \ref{tab:numerical-result2} and
\ref{tab:numerical-result} with those from previous
work of Adelseck-Saghai \cite{Adelseck:1990ch}
(fourth and fifth columns of Table \ref{tab:numerical-result}),
we may conclude that the extracted coupling constants in the 
present work are consistent. However, compared to
the work of Williams {\it et al} \cite{williams} 
and Cheoun {\it et al.} \cite{Cheoun:1996kn}
(sixth and seventh columns of Table \ref{tab:numerical-result}), 
it is found that the extracted couplings of the present 
work are much larger. In view of this, we think that
it is necessary to refit the hadronic coupling constants
in a simultaneous fit to both photo- and electroproduction
data bases. In this case the electroproduction data will
also influence the values of the extracted hadronic coupling
constants. The result is shown in the third column of
Table \ref{tab:numerical-result2} and hereafter is called
PS1. From this result it is 
obvious that the $K^*$ and $K_1$ coupling constants desired
by the electroproduction data are much smaller than those
extracted from the photoproduction data and, surprisingly,
are of the order of those obtained by Williams {\it et al}. 
\cite{williams} and Cheoun {\it et al.} \cite{Cheoun:1996kn}.

\begin{figure}[!h]
  \begin{center}
    \leavevmode
    \epsfig{figure=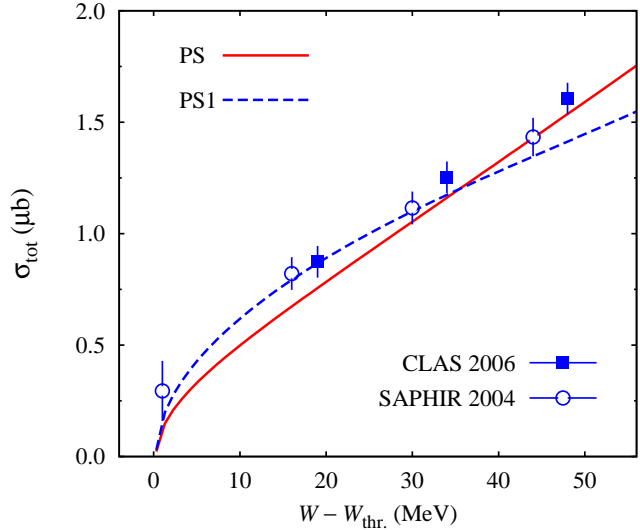,width=90mm}
    \caption{(Color online) Effect of the fitting to both photo- 
      and electroproduction data simultaneously on the photoproduction 
      total cross section. Solid line shows the calculated cross section
      from the PS model, discussed in Section~\ref{sec:result-photo}. 
      Dashed line (PS1) displays the result of fitting to both photo- 
      and electroproduction data, simultaneously.}
   \label{fig:kltotc} 
  \end{center}
\end{figure}

The effect of fitting to both photo- and electroproduction data
simultaneously on the photoproduction total cross section is
displayed in Fig.~\ref{fig:kltotc}. It is easy to understand 
that the effect of electroproduction data is more apparent in
the ``higher'' energy region, since the data exist at $W=1.65$
GeV, about 40 MeV above the threshold. In this region, the
cross section is suppressed due to the size of the electroproduction
data, which are presumably much smaller than those expected by
the photoproduction data. Interestingly, at ``lower'' energies,
the agreement with experimental data is improved after including
the electroproduction data. 

\begin{figure}[!h]
  \begin{center}
    \leavevmode
    \epsfig{figure=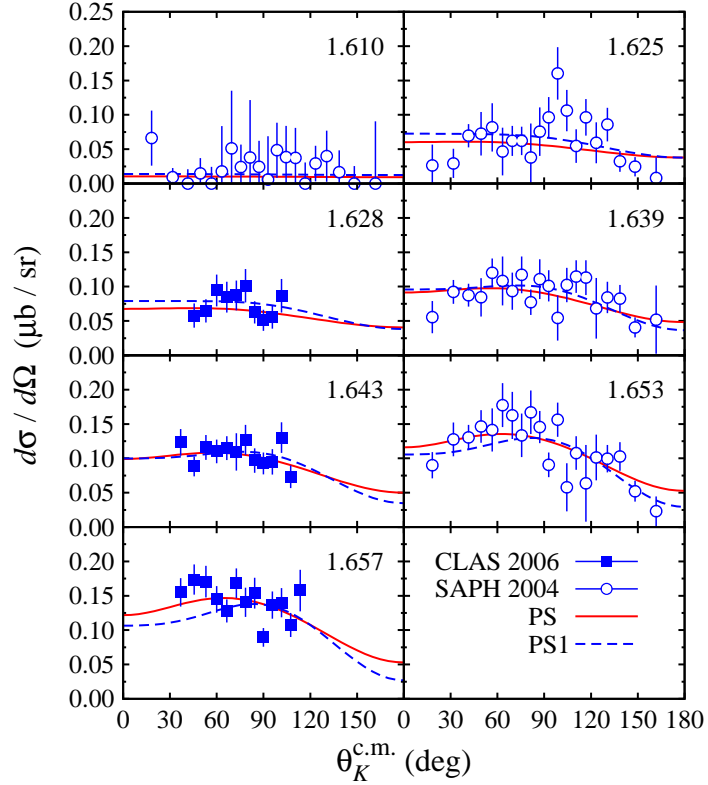,width=100mm}
    \caption{(Color online) As in Fig.~\ref{fig:kltotc}, 
             but for the angular distributions of the differential
             cross section. Experimental data are the same as
             those shown in Fig.~\ref{fig:dif_th}.}
   \label{fig:dif_thc} 
  \end{center}
\end{figure}

To further elucidate the role of electroproduction data on the
suppression of cross sections at $Q^2=0$ we display the angular
distribution of the photoproduction differential cross section 
in Fig.~\ref{fig:dif_thc}. It is found that the suppression
phenomenon occurs only at forward and backward angles
(see panels with $W=1.653$ GeV and $W=1.657$ GeV). This
phenomenon obviously originates from the small values of the
$K^*$ and $K_1$ coupling constants. Unfortunately, experimental
data from CLAS collaboration ($W=1.657$ GeV) are unavailable 
at these kinematical regions. The SAPHIR data at $W=1.653$ GeV
seem to be better in this case. Nevertheless, the corresponding
accuracy is still unable to firmly resolve the difference between
the two calculations, although at the very forward and backward
directions the data seem to favor the small values of the
$K^*$ and $K_1$ coupling constants. Future experimental measurement
at MAMI is expected to settle this problem. 

\begin{figure}[!h]
  \begin{center}
    \leavevmode
    \epsfig{figure=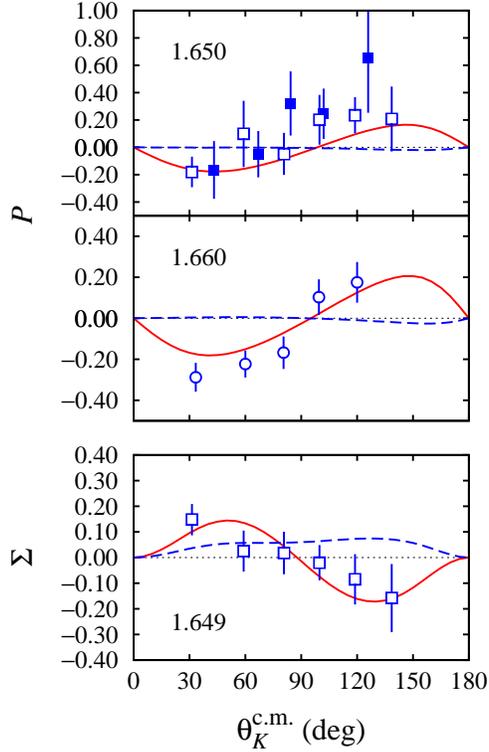,width=70mm}
    \caption{(Color online) As in Fig.~\ref{fig:dif_thc}, 
             but for the angular distributions of the photon
             beam and recoil polarizations. Experimental data 
             are the same as
             those shown in Fig.~\ref{fig:polar}.}
   \label{fig:polarc} 
  \end{center}
\end{figure}

It is also important to note that by comparing the 
calculated photon beam and recoil polarizations 
with experimental data as shown in Fig.~\ref{fig:polarc} 
we can observe the deficiency of the PS1 model. This 
is understandable, since the data shown in this figure
are measured at $W\approx 1.65$ GeV, an energy region where
predictions of the PS1 model for photoproduction 
are no longer reliable, due to the strong
influence of the electroproduction data
(see Fig.~\ref{fig:kltotc}). Furthermore, 
the number of the polarization data
in the fitting database cannot compete with that of the cross 
section data. Increasing the number and improving the accuracy
of these data near the threshold 
or using the weighting factor in the fitting process
could be expected to improve this situation. 

\begin{figure}[!t]
  \begin{center}
    \leavevmode
    \epsfig{figure=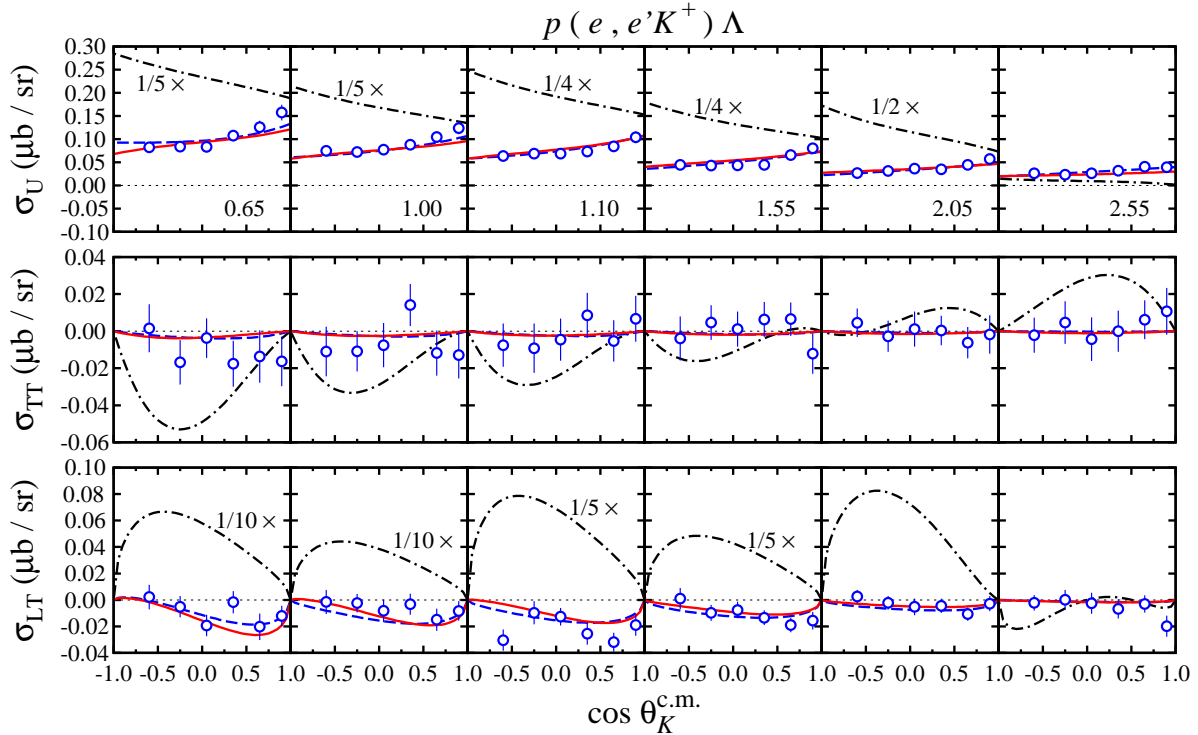,width=160mm}
    \caption{(Color online) Separated differential cross sections for 
      kaon electroproduction $e+p\to e'+K^++\Lambda$ as a function
      of the kaon scattering angles at $W=1.65$ GeV and
      for different values of $Q^2$ (the values are shown in the top panels). 
      Experimental data are from the CLAS collaboration
      \cite{Ambrozewicz:2006zj}. 
      Solid lines are due to the PS model, dashed lines show
      the result of the PS1 model, whereas dash-dotted lines
      display the predictions of Kaon-Maid.
      Note that predictions of Kaon-Maid in some panels have 
      been renormalized by certain factors in order to fit on the scale.}
   \label{fig:ambroz} 
  \end{center}
\end{figure}

\begin{figure}[!h]
  \begin{center}
    \leavevmode
    \epsfig{figure=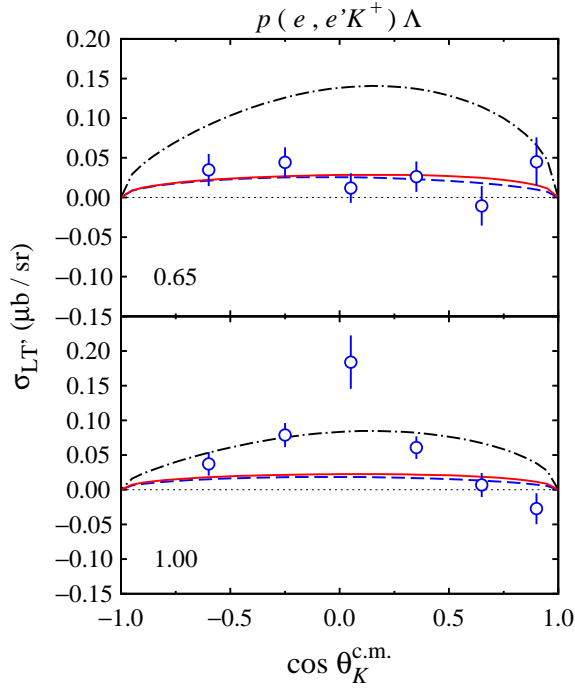,width=80mm}
    \caption{(Color online) Polarized structure 
      function $d\sigma_{\rm LT'}/{d\Omega}$
      of the $e+p\to e'+K^++\Lambda$ channel as a function of the 
      kaon scattering at $W=1.65$ GeV with 
      $Q^2=0.65$ GeV$^2$ (upper panels) and $Q^2=1.00$ GeV$^2$ 
      (lower panels). Notation of the curves is as in 
      Fig.~\ref{fig:ambroz}. Experimental data are from 
      Ref. \cite{Nasseripour}.}
   \label{fig:nasser} 
  \end{center}
\end{figure}

A comparison between separated differential cross sections 
$\sigma_{\rm U}$, $\sigma_{\rm TT}$, and $\sigma_{\rm LT}$
calculated from the PS, PS1, and Kaon-Maid models with experimental
data is shown in Fig.~\ref{fig:ambroz}. In this case
it is clear that the PS and PS1 can nicely reproduce 
the data and  their difference is hardly seen. 
The same result is also shown in
case of the polarized structure function $\sigma_{\rm LT'}$, 
as displayed in Fig.~\ref{fig:nasser}. However, we notice
that Kaon-Maid is unable to reproduce the available data
in all cases and, in fact, it shows a backward 
peaking behavior for $\sigma_{\rm U}$,
in contrast to the result of experimental measurement. We
suspect that this behavior originates from the large
contributions of the $K^*$ and $K_1$ intermediate states
(see the fourth column of Table~\ref{tab:numerical-result}).
In the PS or PS1 model their contributions are significantly
suppressed by means of the exponential form factors or 
small values of $K^*$ and $K_1$ coupling constants, 
respectively.

It is also important to note that both PS and PV model seem to 
choose the zero $\sigma_{\rm TT}$, a conclusion that
has been also drawn in Ref.~\cite{Ambrozewicz:2006zj}. However,
we notice that this is 
in contrast to the prediction of Kaon-Maid.

\begin{figure}[!h]
  \begin{center}
    \leavevmode
    \epsfig{figure=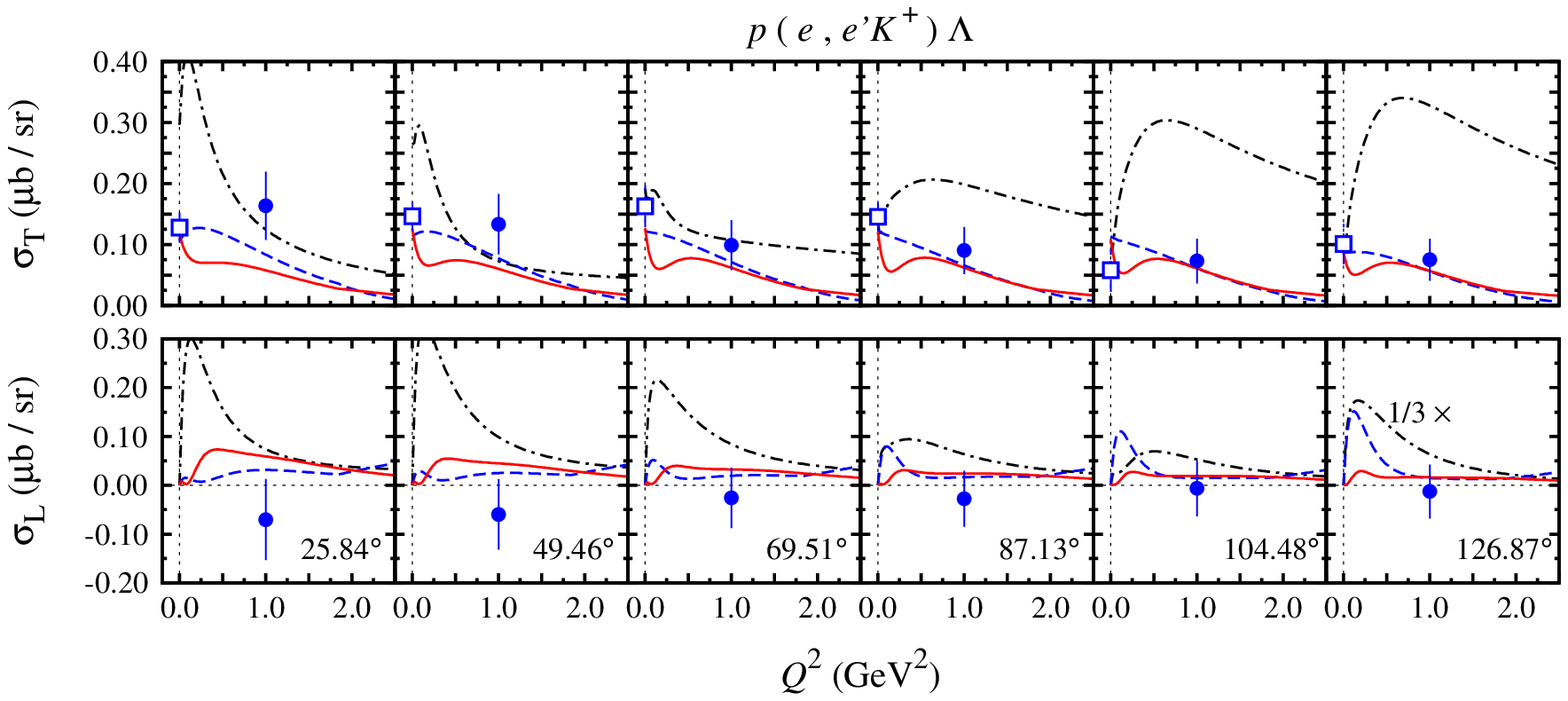,width=160mm}
    \caption{(Color online) Separated transverse (upper panels) and 
      longitudinal (lower panels) differential cross sections 
      of the $e+p\to e'+K^++\Lambda$ channel as a function of the 
      virtual photon momentum transfer 
      $Q^2$  at $W=1.65$ GeV and different $\theta_K^{\rm c.m.}$
      (shown in lower panels). 
      Notation of the curves is as in Fig.~\ref{fig:ambroz}. 
      Experimental data of electroproduction 
      at $Q^2=1.0$ GeV$^2$ (solid circles) 
      are from Ref. \cite{Ambrozewicz:2006zj}, whereas data 
      of photoproduction (open squares) 
      are due to the SAPHIR collaboration \cite{Glander:2003jw}.
      Note that predictions of Kaon-Maid have been renormalized 
      by a factor 1/3 in the lower panels 
      in order to fit on the scale. The electroproduction data
      shown in this figure are not used in our calculation.}
   \label{fig:sl_st} 
  \end{center}
\end{figure}

The $Q^2$ evolution of the longitudinal cross section $\sigma_{\rm L}$ is 
of special interest, especially for estimating the possibility of 
extracting kaon form factor $F_K(Q^2)$. In the latest measurement 
at JLab there has been an attempt to separate $\sigma_{\rm L}$ and
 $\sigma_{\rm T}$ from  $\sigma_{\rm U}$. Except at $W=1.95$ GeV
results of both the Rosenbluth and the simultaneous $\epsilon-\Phi$ 
fit techniques are plagued with the ``nonphysical'' 
(negative) longitudinal cross section $\sigma_{\rm L}$. 
Nevertheless, it was concluded that within the
combined systematical and statistical uncertainties the longitudinal
cross sections are believed to be zero \cite{Ambrozewicz:2006zj}. 

The result of our PS and PS1 models are compared with the 
prediction of Kaon-Maid and very limited experimental data 
in Fig. \ref{fig:sl_st}. The agreement of the 
PS and PS1 models with the photoproduction data for transverse
cross section at $Q^2=0$ is certainly not surprising. However,
the discrepancies between experimental data and model calculations
at $Q^2=1$ GeV$^2$ still show the consistency of the presented models.
It should be remembered that both PS and PS1 models were fitted to the
same, but unseparated data $\sigma_{\rm U}$. Therefore, as shown
in upper and lower panels with the same $\theta_K^{\rm c.m.}$ in 
Fig. \ref{fig:sl_st}, the underpredicted $\sigma_{\rm T}$ will always
be compensated by the overpredicted  $\sigma_{\rm L}$. This result
clearly indicates that in order to avoid the ``nonphysical'' 
separated longitudinal cross section $\sigma_{\rm L}$, 
a model-dependent extraction of  $\sigma_{\rm L}$ and
$\sigma_{\rm T}$ would be strongly recommended. 
Obviously, this could be performed only if 
we had a reliable isobar model that can describe all 
available data.

From the smallness of the longitudinal cross sections shown in
Fig.~\ref{fig:sl_st}, it is obvious that the extraction of $K^+$
form factor is difficult near the production threshold. 
Moreover, we have numerically found that
the longitudinal cross sections shown in the lower panels of 
Fig. \ref{fig:sl_st} are dominated by the $S_{11}(1650)$ contribution.
This is understandable since these cross sections are calculated at
$W=1.65$ GeV, precisely at the pole position. In 
Ref.~\cite{Ambrozewicz:2006zj} it is shown that the ``physical'' 
(positive) longitudinal cross section data are only found
at higher $W$, i.e. 1.95 GeV. 
From this fact, the extraction of the kaon electromagnetic 
form factor is naturally recommended at higher energies.
Especially at $W$ above 2 GeV, where the 
contribution of nucleon resonances becomes less significant
than the background terms.
Furthermore, the extraction is best performed at small $|t|$,
i.e., forward angles, where the contribution of $t$-channel
is maximum.

For completeness we would like also to mention that we have
investigated also kaon electroproduction in the case of PV
coupling and found no substantial difference from the result
of PS coupling.

\section{Summary and Conclusions}
\label{sec:conclusion}

We have investigated kaon photo- and electroproduction
off a proton near the production threshold by means of 
an isobar model. The background amplitude is constructed
from the appropriate Feynman diagrams, whereas the resonance
amplitude is calculated by using the multipole formalism.
In contrast to the results of the previous works that
utilize isobar model or chiral perturbation theory 
it is found that both PS and
PV models in the present work can nicely describe the available 
photoproduction data up to $W=50$ MeV above the threshold. 
Since in this work the
values of the main coupling constants $g_{K\Lambda N}$ 
and $g_{K\Sigma N}$ are fixed to the SU(3) symmetry prediction,
the obtained $\chi^2$ in the PV model is larger than
in the case of the PS model. This originates from the fact that
for the same coupling constants, the background amplitudes of
the PV model is larger than that of the PS model.
The $K^*$ and $K_1$ coupling
constants are found to be consistent with the previous works.
The $\Lambda$ resonance $S_{01}(1800)$ is found to play
an important role in improving the agreement
of the model calculation with experimental data, 
especially in the case of photon and recoil polarizations. 

It is found that the $K^*$ and $K_1$ hadronic coupling constants 
extracted from photoproduction data lead to very large 
electroproduction differential cross sections, and therefore 
could not be used
unless special form factors that strongly suppress their
contributions were introduced or the hadronic coupling constants were
refitted to both photo- and electroproduction databases,
simultaneously. In the latter, the model reliability at
lower energies is improved, whereas at higher energies the
calculated photoproduction observables significantly deviate
from experimental measurement due to the strong influence of 
electroproduction data. 
From the size of longitudinal cross sections $\sigma_{\rm L}$
predicted by both PS and PS1 models we may
conclude that investigation of the kaon electromagnetic form
factor, that strongly relies on  $\sigma_{\rm L}$, 
is difficult to perform near the threshold region.
More accurate experimental data at this energy region would 
certainly help to clarify some uncertainties in the present work.

\section*{Acknowledgment}
The author acknowledges supports from the University of Indonesia
and the Competence Grant of the Indonesian 
Ministry of National Education.

\end{document}